%% file: Jacobi.tex
\documentclass[12pt]{iopart}

\usepackage{iopams}
\usepackage{epsfig,bm}
\usepackage{graphicx}
\usepackage{subfigure}
\usepackage{amssymb,psfrag,textcomp}
\usepackage{amsthm,upgreek}

\newcommand{\vD}{\mathcal{D}}
\newcommand{\xD}{\Delta}
\newcommand{\zone}{\zeta_1}
\newcommand{\ztwo}{\zeta_2}
\newcommand{\aone}{\alpha_1}
\newcommand{\atwo}{\alpha_2}
\newcommand{\vJ}{\mathcal{J}}

\newcommand{\vI}{\mathcal{I}}
\newcommand{\vW}{\mathcal{W}}
\newcommand{\vX}{\mathcal{X}}
\newcommand{\vO}{\cal{O}}

\begin{document}

\title{Spectral Properties of the Jacobi Ensembles via the Coulomb Gas approach}

\author{Huda Mohd Ramli, Eytan Katzav and Isaac P\'erez Castillo}

\address{Department of Mathematics, King's College London, Strand, London WC2R 2LS, United Kingdom}
\ead{huda.ramli@live.com}
\ead{eytan.katzav@kcl.ac.uk}
\ead{isaac.perez{\_}castillo@kcl.ac.uk}
\begin{abstract}
Using the Coulomb gas method and standard methods of statistical
physics, we compute analytically the joint cumulative probability
distribution of the extreme eigenvalues of the Jacobi-MANOVA
ensemble of random matrices, in the limit of large matrices. This
allows us to derive the rate functions for the large fluctuations to
the left and the right of the expected values of the smallest and
largest eigenvalues analytically. Our findings are compared with some available known exact
results as well as with numerical simulations finding good agreement.

\end{abstract}

\maketitle

\section{Introduction}

Random Matrix Theory (RMT), the study of ensembles of matrices with
random entries, has been actively explored since the pioneering
works of Wigner \cite{wigner} and Dyson \cite{dyson}, sixty years
ago, and it still remains a very active and challenging area of
research, with countless applications in mathematical physics,
statistical mechanics and beyond. The three classical ensembles of
RMT are the Gaussian (or Hermite), Wishart (or Laguerre) and Jacobi
(or MANOVA) ensembles. The Gaussian ensemble appear in physics and
are obtained as the eigenvalue distribution of a symmetric matrix
with Gaussian entries. Both the Wishart and Jacobi ensembles are
very common in statistics - where Wishart corresponding to the
distribution of covariance matrices, and the Jacobi ensemble
motivated by multivariate analysis of variance (MANOVA), see for
example Muirhead \cite{muirhead} for real valued matrices.

Let us first look at the Wishart ensemble which comprises of the
product matrices $\vW = \vX^\dag \vX$ of an $M \times N$ matrix
$\vX$ having its elements drawn independently from a Gaussian
distribution, $P[\vX] \propto \exp\bigl[-\frac{\beta}{2}
Tr(\vX^\dag \vX)\bigr]$. The Dyson indices $\beta = 1,2,4$
correspond to real, complex and quaternion entries of $\vX$
respectively. In essence, Wishart matrices are random covariance
matrices, constructed from random data matrices $\vX$ (typically real
entries in statistical application). By construction, such matrices are positive
definite. Originally introduced by Wishart in 1928, these Wishart
matrices have not only been extensively applicable in multivariate
statistics, but have also been proved to be useful in areas of
nuclear physics \cite{nuclearphy}, quantum chromodynamics
\cite{quantum} and statistical physics \cite{statphys}. Large
deviation properties have also been widely investigated successfully
in \cite{maxwishart,wishartgaussian}.

The Jacobi/MANOVA matrix $\vJ$ is an $N \times N$ matrix that is a
combination of two Wishart matrices, $\vW(N,M_1)$ and $\vW(N,M_2)$ in
the form:
\begin{equation*}
\vJ (N,M_1,M_2) = \frac {\vW(N,M_1)}{\vW(N,M_1)+ \vW(N,M_2)} \, ,
\end{equation*}
where $N$, $M_1$ and $M_2$ are the dimensions of the matrices Note that the definition of Jacobi/MANOVA matrix implies that $M_1,M_2 \ge N$. As will be seen below we will be interested in the limit of large matrices, i.e. $N, M_1, M_2 \gg 1$, such that the ratios $c_1 = N/M_1$ and $c_2 = N/M_2$ are kept constant. Therefore, we will use the notation $\vJ (c_1,c_2)$ from now on, with $c_1,c_2 \le 1$ (similar to the notation used for the Wishart matrices $\vW(c)$). The case $c_1=c_2=1$ (or $M_1=M_2=N$) corresponds to square matrices and the case $c_i-1={\vO}(N^{-1})$ (or $M_i-N={\vO}(1)$) is referred to as "almost square matrices". Last, note that by definition, the eigenvalues of $\vJ$ are positive and bounded from above by $1$, namely by definition $\lambda_{min} \ge 0$ and $\lambda_{max} \le 1$.

The deviation of the extreme eigenvalues of Jacobi ensembles from their expectation values plays a significant role in multivariate statistics, such as the Canonical Correlation Analysis (CCA). Suppose we have two data matrices, one $N\times M_1, X=[x_1 x_2 \dots x_{M_1}]$ and the other $N\times M_2, Y=[y_1 y_2 \dots y_{M_2}]$. The idea behind CCA is to seek linear combinations of $a^Tx$ and $b^Ty$ that are most highly correlated.
Formally, this is done by determining the largest "latent root" of the generalized eigenvalue problem: $B-r^2(A+B)$, where $B=X^TPX, A=X^TP^{\bot}$ and $P$ is the $N\times N$ orthogonal projection matrix $Y(Y^TY)^{-1}Y^T$ and $P^{\bot}=I-P$ is its complement \cite{johnstone}. Another application of the convergence is in the standard generalization of the linear regression model to allow for multivariate responses, which is discussed in greater detail in
Johnstone(2009) \cite{johnstone2}. Other interesting Jacobi applications include Quantum conductance in mesoscopic physics \cite{vivothesis} and the multivariate F distribution \cite{dettenagel}.

During the years, it has been recognized that the real, complex and
quaternion Jacobi random matrices are special cases of a larger
family of random matrices now known as the $\beta$-Jacobi ensemble,
and which was discovered by \cite{Suttonthesis,Lippert}. In that
context, the real, complex and quaternion Jacobi matrices correspond
to the Dyson index with $\beta = 1, 2$ and $4$ respectively, and
they are special in that only for these cases there exists a matrix
model that is not sparse. Many expressions describing statistical properties of these ensembles have been successfully derived, such as formulas for the moments for any matrix size $N$ \cite{Livan}, as well as a closed form expression for the spectral density $\rho_N(\lambda)$ for large $N$ \cite{johnstone,Forrester,Ledoux,Collins} (essentially a generalization of the Marecenko-Pastur law for the Jacobi ensemble), namely
\begin{equation}
\rho_N(x) = \frac{c_1+c_2}{2\pi}\frac{\sqrt{(x-\zeta_-)(\zeta_+-x)}}{x(1-x)} 1_{x \in [\zeta_-,\zeta_+]} \, ,
\label{eq:JMP}
\end{equation}
with so called natural boundaries
\begin{equation}
\zeta_{\pm}=\frac{(c_1-1)(c_1+c_2)+2\left[c_2 \pm \sqrt{c_1 c_2(c_1+c_2-1)}\right]}{(c_1+c_2)^2} \, .
\label{eq:zetas0}
\end{equation}
The indicator $1_{x \in D}$ takes the value $1$ if $x \in D$, and zero otherwise.
As can be seen this spectral density has a compact support $[\zeta_-,\zeta_+]$ suggesting that as $N$ becomes large the eigenvalues are bounded in a non-trivial way. Also note that in general $\zeta_- \ge 0$ and $\zeta_+ \le 1$, which reflect the absolute constraints $\lambda_{min} \ge 0$ and $\lambda_{max} \le 1$ mentioned above. Interestingly, for square and almost-square matrices $\zeta_-=0$ and $\zeta_+ = 1$, and so the natural boundaries merge with the hard walls at $0$ and $1$, which makes this limit somewhat different from the generic non-square case.

More detailed spectral information about the Jacobi ensemble has been achieved in \cite{Jiang09} where the expectation values of all the eigenvalues have been derived (again in the large $N$ limit), including the expectations for the largest and smallest eigenvalues. Further more, the cumulative distribution functions (CDFs) for the extremal eigenvalues in terms of the hypergeometric function of a matrix argument have been derived in \cite{beta-jacobi,Dumitriu} for any $\beta$, and a simpler determinantal expression for the distribution of smallest eigenvalue has been derived in \cite{Collins} for the classical ensembles. The determinantal representation lends itself to proving that the {\bf typical} deviations of the largest/smallest eigenvalue from its expectation can be describes using the the celebrated Tracy--Widom (TW) distribution \cite{TW} as long as $c_1,c_2<1$ (i.e., not for square or almost-square matrices), as has been demonstrated by Johnstone in \cite{johnstone}. Following this, the typical fluctuations of $\lambda_{max}$ can be expressed as (see \ref{app:TW} for more details)
\begin{equation*}
\lambda_{\max} \sim \zeta_+ + N^{-2/3} \frac{\left[ \zeta_+ (1-\zeta_+) \right]^{2/3}}{[c_1 c_2(c_1+c_2-1)]^{1/6}}  Z_\beta + \Or \left( N^{-4/3} \right) \, ,
\end{equation*}
where $Z_\beta$ follows a TW distribution $F_\beta$ \cite{johnstone}. A similar expression can be written for $\lambda_{min}$ as well (see \ref{app:TW}).This result is consistent with an exact inequality for the deviations around $\lambda_{max}$ derived in \cite{Ledoux}, which also confirms the non-obvious scaling with $N$. 

As mentioned above, the distributions of the smallest and largest eigenvalues have numerous applications in multivariate statistics, but they are in practice difficult to calculate exactly, except for fairly small matrices. Moreover, knowledge of large deviations of the extreme eigenvalues is mostly unexplored. 
The main purpose of this paper is to provide exact large $N$ analytical results for these large fluctuations of the extreme eigenvalues $\lambda_{max}$ and $\lambda_{min}$ from their mean values, which would be as simple as possible and practical.
This is achieved by the use of two ingredients: the customary Coulomb gas mapping, which interprets the eigenvalues of a random matrix as charged interacting particle (see \cite{deanmajumdar} and also \cite{chenmanning} specifically for the Wishart-Laguerre case), and standard functional integration methods of statistical physics. This technique has been recently used to obtain analytically the large negative fluctuations of the maximum eigenvalue \cite{maxwishart} (as well as for the minimal eigenvalue \cite{WLensemble}) for the Wishart ensemble. Here we adopt this method for the Jacobi ensemble. We show that the logarithm of the probability of large fluctuations to the right/left of the mean value of $\lambda_{min}$/$\lambda_{max}$, for large $N$, can be written asymptotically as
\begin{eqnarray}
\log P_N^{(\min)}(\zone) \sim -\beta N^2 \Phi_+^{(\min)}(\zone - \zeta_-), \quad \zone \geq \zeta_- \label{eq:minp} \\
\log P_N^{(\max)}(\ztwo) \sim -\beta N^2 \Phi_-^{(\max)}(\zeta_+ - \ztwo), \quad \ztwo \leq \zeta_+ \label{eq:maxm}
\end{eqnarray}
where $\Phi_{\pm}(x)$ are right/left rate functions (also known as the large deviation functions) that are derived explicitly with $x$ being the main argument of the function. These expressions are valid for any large $N$ size of Jacobi ensembles and any value of $\beta$ including $\beta=1$ (real), $2$ (complex) and $4$ (quaternions) as special cases.

Strictly speaking the Coulomb-gas method combined with the saddle-point integration can only yield the two rate functions just mentioned. However, thanks to an insightful physical argument by Majumdar and Vergassola developed for Wishart matrices \cite{wishartgaussian} we will  be able to derive the other two rate functions, as will be explained later.

Some words about the organization of this paper. In \sref{sec:two}, we set up notations and provide some mathematical preliminaries and recall some known results for the Jacobi ensembles which serve to set up our notations for the rest of the paper. In \sref{sec:method}, we outline the Coulomb gas method followed by the saddle-point approximation. The Hilbert transformation is then used to reach a general expression of the Jacobi eigenvalue spectral density. In \sref{sec:wall}, we derive the probability distributions explicitly for four different cases of constraints around the boundaries of the spectral density. Using the Majumdar-Vergassola approach \cite{wishartgaussian} in \sref{sec:othersides}, we find the other rate functions that could not be captured by the saddle-point method applied before. In \sref{sec:num} the analytical predictions are compared to numerical simulations. \Sref{sec:concl} concludes the paper with a summary and discussion. Finally the technical derivations of integrals applied in \sref{sec:method} are detailed in the appendices.

\section{Jacobi Ensembles: some preliminaries}
\label{sec:two}
As explained in the Introduction, we focus on the Jacobi ensemble, with $\vJ (c_1,c_2)= \frac {\vW(c_1)}{\vW(c_1)+ \vW(c_2)}$, where $c_i =\frac{N}{M_i}$ ($i=1,2$), and $\vW = \vX^\dag \vX$ of an ($M \times N$) random Gaussian matrix $\vX$. For $M_i \ge N$ $(i=1,2)$, the joint probability distribution function (jpdf) of the eigenvalues of $\vJ(c_1,c_2)$, simultaneously derived in 1939 by Fisher, Girshick, Hsu, Mood and Roy (see Muirhead~\cite{muirhead} page 112), can be written in closed form as
\begin{equation}
\hspace{-1cm} P_N(\lambda_1, \dots ,\lambda_N) \propto \prod_{i=1}^N{\lambda_i}^{\frac{\beta}{2}(M_1-N+1)-1} {(1-\lambda_i)}^{\frac{\beta}{2}(M_2-N+1)-1} \prod_{i < j} {|\lambda_i-\lambda_j|}^\beta \, .
\label{eq:jpd}
\end{equation}
Note that the jpdf in \Eref{eq:jpd} above has a symmetry under the exchange of $(M_1 \leftrightarrow M_2)$ and $(\lambda_i\leftrightarrow 1-\lambda_i)$. This symmetric feature of the distribution has important and useful consequences that will be discussed later. In short, it entails a certain equivalence between the distributions of the smallest and largest eigenvalues for the Jacobi ensemble.

We would like to compute the probability $P_N(\zone,\ztwo)$ that all eigenvalues are in the interval $[\zone,\ztwo]$ where $0 \leq \zone\leq \ztwo\leq 1$ using the Coulomb gas method (note again that all eigenvalues obey $0 \le \lambda_i \le 1$ by construction of the Jacobi matrix). Evidently, $P_N(\zone,\ztwo)$ is also the cumulative probability that the minimum eigenvalue $\lambda_{min}$ is greater than or equal to $\zone$, and at the same time the maximum eigenvalue $\lambda_{max}$ is less than or equal to $\ztwo$, i.e.,
\begin{eqnarray}
P_N(\zone,\ztwo)&=& {\rm Prob}[\zone\leq\lambda_1\leq \ztwo,\zone\leq\lambda_2\leq \ztwo, \dots,\zone\leq\lambda_N\leq \ztwo] \nonumber \\
&=& {\rm Prob}[\lambda_{min} \geq \zone, \lambda_{max} \leq \ztwo] \, .
\label{eq:jpde}
\end{eqnarray}
In other words, $P_N(\zone,\ztwo)$ provides the joint probability distribution of the minimum and the maximum eigenvalue, which can be therefore expressed as
\begin{equation}
P_N(\zone,\ztwo)= \int_{\zone}^{\ztwo} \dots \int_{\zone}^{\ztwo} P_N(\lambda_1, \dots ,\lambda_N) d\lambda_1 \dots d\lambda_N \, .
\label{eq:jpde2}
\end{equation}

\section{Analytics: The Coulomb Gas Approach}
\label{sec:method}
Our starting point in the analytical treatment is the joint distribution of eigenvalues of the Jacobi ensemble (\ref{eq:jpd}) which can be presented as
\begin{equation}
P_N(\lambda_1, \dots ,\lambda_N) = \frac{1}{Z_0} \exp{\left(-\frac{\beta}{2}F(\bm \lambda)\right)} \, ,
\label{eq:partfcn}
\end{equation}
where $Z_0$ is a normalization constant, and the function $F(\bm \lambda)$ is defined as
\begin{eqnarray}
\hspace{-2cm} F({\bm \lambda}) &=& -\left(M_1-N+1-\frac{2}{\beta} \right) \sum_{i=1}^N \log{\lambda_i} - \left(M_2-N+1-\frac{2}{\beta}\right)\sum_{i=1}^N\log(1-\lambda_i)\nonumber \\
 \hspace{-2cm}&& -\sum_{i=1}^N \sum_{j \neq i} \log{|\lambda_i-\lambda_j|} \, .
\label{eq:freeenergy}
\end{eqnarray}

The quantity we are interested in is the joint probability of the extreme eigenvalues $P_N(\zone,\ztwo)$ given as a multiple integral by Eq.~(\ref{eq:jpde2}), which can be written as a ratio of two so-called (see below) partition functions
\begin{equation}
\hspace{-1cm}P_N(\zone,\ztwo)= \frac{1}{Z_0}\int_{\zone}^{\ztwo} d\lambda_1 \dots \int_{\zone}^{\ztwo} d\lambda_N \exp\left(-\frac{\beta}{2} F(\bm \lambda) \right) = \frac{Z(\zone,\ztwo)}{Z(0,1)} \, ,
\label{eq:jointp}
\end{equation}
with
\begin{eqnarray*}
Z(\zone,\ztwo)&=& \int_{\zone}^{\ztwo} d\lambda_1 \dots \int_{\zone}^{\ztwo} d\lambda_N \exp\left(-\frac{\beta}{2}F(\bm \lambda)\right) \,,
\end{eqnarray*}
with $Z_0=Z(0,1)$. Written in this form, $F(\bm \lambda)$ can be seen as the free energy of a two-dimensional (2D) Coulomb gas of charged particles (which repel each-other) constricted onto the real line with a logarithmic-logarithmic external potential. $Z(0,1)$ is simply a normalization constant, representing the partition function of a free or unconstrained Coulomb gas over $\lambda\in[0,1]$, while $Z(\zone,\ztwo)$ is the partition function of the same Coulomb gas but confined inside the segment $\lambda\in[\zone,\ztwo]$.

The idea here is to evaluate the partition function, $Z(\zone,\ztwo)$ in the large $N$ limit using the saddle-point method.

Before proceeding further, we must first introduce the constrained charge density, which describes the density profile of the gas in the presence of two hard walls constraining the gas particles inside $[\zone,\ztwo]$ (from a probabilistic point of view this is simply the conditioned spectral density)
\begin{equation*}
f(x) \equiv \rho(\lambda=x|\zone \le x \le \ztwo)=\frac{1}{N} \sum_{i=1}^N \delta(x-\lambda_i) \Theta(x-\zone) \Theta(\ztwo-x) \, .
\end{equation*}
This allows us to write $Z(\zone,\ztwo)$ as a path integral over $f(x)$ and its Lagrange multiplier $\hat{f}(x)$ that enforces its definition \cite{deanmajumdar,maxwishart,WLensemble}.
$Z(\zone,\ztwo)$ now takes the form
\begin{equation*}
Z(\zone,\ztwo) = \int \lbrace\vD f\rbrace e^{-\frac{\beta N^2}{2}S[f(x)]} \, ,
\end{equation*}
where $S[f(x)]$ is the action, given by
\begin{eqnarray*}
 \hspace{-2cm} S[f(x)] = -\int_{\zone}^{\ztwo} dx \int_{\zone}^{\ztwo}\, dx' f(x)f(x')\log|x-x'| - \left( \aone+\frac{\beta-2}{\beta N} \right)\int_{\zone}^{\ztwo}\, dx f(x)\log x \nonumber \\
-\left(\atwo+\frac{\beta-2}{\beta N} \right)\int_{\zone}^{\ztwo}dx f(x)\log (1-x) \nonumber \\
+\frac{1}{N}\int_{\zone}^{\ztwo}dx f(x)\log f(x) + A\left(\int_{\zone}^{\ztwo} dx f(x)-1\right) \, ,
\end{eqnarray*}
where $A$ in the last term is the Lagrange multiplier enforcing the normalization of $f(x)$ and we have defined the parameters $\alpha_i=\frac{M_i}{N}-1$, $i=1,2$ for convenience.

We are interested in large Jacobi matrices, and consider the large $N$ limit in this calculation. Thus we neglect terms of order $\vO\)\((N^{-1})$ in the action $S[f(x)]$, and will consider the leading order terms in the action, namely
\begin{eqnarray}
\hspace{-2cm} S[f(x);\zone,\ztwo] = -\int_{\zone}^{\ztwo} dx \int_{\zone}^{\ztwo} dx' f(x)f(x')\log|x-x'| - \aone\int_{\zone}^{\ztwo}dx f(x)\log x
\nonumber \\
\qquad -\atwo\int_{\zone}^{\ztwo}dx f(x)\log (1-x) +A\left(\int_{\zone}^{\ztwo} dx f(x)-1\right) \, ,
\label{eq:action}
\end{eqnarray}
which yields
\begin{equation}
Z(\zone,\ztwo) \propto \exp\left(-\frac{\beta N^2}{2}S[f(x);\zone,\ztwo]\right) \, .
\label{eq:pfzeta}
\end{equation}
For large $N$, we can evaluate the leading contribution to the action \eref{eq:action} using the saddle-point equation which reads $\delta S[f(x)] / \delta f(x) =0$
\begin{equation}
-2\int_{\zone}^{\ztwo} dx' f_*(x')\log|x-x'| -\aone\log x -\atwo\log(1-x) + A = 0 \, ,
\label{eq:saddle}
\end{equation}
where $f_*(x)$ if the value of $f(x)$ at the saddle point.
Differentiating Eq.~(\ref{eq:saddle}) once with respect to $x$, we obtain
\begin{equation}
-\frac{\aone}{2x} +\frac{\atwo}{2(1-x)} = P\int_{\zone}^{\ztwo} dx'\frac{f_*(x')}{x-x'} \, ,
\label{eq:integral}
\end{equation}
with $x\in [\zone,\ztwo]$ and $P$ denotes the Cauchy Principal Value.\\
\Eref{eq:integral} is a Tricomi integral equation whose solution involves the Hilbert transform. The general solution of equations of the type
\begin{equation*}
\frac{1}{\pi}\int_a^b \frac{\varphi(t)}{t-x}dt=\psi(x), \qquad a\leq x\leq b
\end{equation*}
is given by Tricomi's theorem \cite{tricomi}:
\begin{equation*}
\varphi(x) =\frac{1}{\pi^2} \frac{1}{\sqrt{(x-a)(b-x)}} \left(P\int_a^b \frac{\sqrt{(t-a)(b-t)}}{t-x} \psi(t)dt + C\right) \, ,
\end{equation*}
where $C$ is an arbitrary constant. In the particular case discussed here (\ref{eq:integral}), we have
\begin{equation}
\fl f_*(x) =\frac{1}{\pi^2}\frac{1}{\sqrt{(x-\zone)(\ztwo-x)}}
\left(P\int_{\zone}^{\ztwo}\frac{\sqrt{(t-\zone)(\ztwo-t)}}{t-x}
\left(\frac{\atwo}{2(1-t)}-\frac{\aone}{2t}\right)dt + C\right) \, .
\label{eq:tricomi}
\end{equation}
After various changes of variables, which are described in detail in \ref{app:transformations}, we obtain the final and general expression of the limiting spectral density:
\begin{eqnarray}
 f_*(x;\zone,\ztwo)&=&\frac{1}{2\pi\sqrt{(x-\zone)(\ztwo-x)}}\Bigg[\aone\left(1-\frac{\sqrt{\zone\ztwo}}{x}\right)\nonumber \\
&+&\atwo\left(1-\frac{\sqrt{(1-\zone)(1-\ztwo)}}{1-x}\right)+2\Bigg] 1_{x \in [\zone,\ztwo]} \, .
\label{eq:final}
\end{eqnarray}

Now we can substitute Eq.~\eref{eq:final} back into the saddle-point equation \eref{eq:saddle} to find the value of the Lagrange multiplier $A$ which is explicitly derived in \ref{app:case1}. It turns out that the integrals involved can be analytically solved using a change of variables, $x=y\xD+\zone$ where $\xD=\ztwo-\zone$, and we eventually arrive at the following expression for the action evaluated at the saddle point:
\begin{eqnarray}
\fl S[f_*(x);\zone,\ztwo] &=& -(\aone+\atwo+2) \left(\aone\log\frac{\sqrt{\zone}+\sqrt{\ztwo}}{2}
+\atwo\log\frac{\sqrt{1-\zone}+\sqrt{1-\ztwo}}{2} \right) \nonumber \\
\fl &+& \frac{{\aone}^2}{2} \log \sqrt{\zone\ztwo}+\frac{{\atwo}^2}{2} \log \sqrt{(1-\zone)(1-\ztwo)} \nonumber \\
\fl &+&\aone\atwo \log\frac{\sqrt{\ztwo(1-\zone)}+\sqrt{\zone(1-\ztwo)}}{2}-\log\frac{\ztwo-\zone}{4} \, .
\label{eq:finalaction}
\end{eqnarray}

As a consequence of the symmetry highlighted previously in the joint probability distribution (\ref{eq:jpd}), the action (\ref{eq:finalaction}) too has symmetry properties. This can be seen clearly if we exchange $(\aone\leftrightarrow\atwo)$ and $(\zeta_i\leftrightarrow 1-\zeta_i)$, in which case the expression is mapped onto itself. This observation has two important applications. First, it serves as a reliable check for our formulae (including our spectral density in (\ref{eq:final})). Second, it also allows us to predict that the expression for the probability of the minimum eigenvalue, $P_N^{(min)}(\zeta)$ with $(\aone,\atwo)$ is equivalent to that of the maximum eigenvalue, $P_N^{(max)}(1-\zeta)$ with $(\atwo,\aone)$.

\section{The Constrained Spectral Density}
\label{sec:wall}
We must note that the solution in \eref{eq:final} gives the general average density of eigenvalues in the limit of large $N$ for the ensemble of Jacobi matrices whose rescaled eigenvalues are restricted to $x\in [\zone,\ztwo]$. From past experience with the Jacobi ensemble we expect the solution to the integral equation to yield a spectral density with a compact support, even when the distribution is not constrained at all (this is similar to the Marcenko-Pastur law in the Wishart ensemble - see \cite{maxwishart,WLensemble}). These natural boundaries $x\in [\zeta_-,\zeta_+]$ which defines the unconstrained normalized eigenvalue distribution i.e. $f_*(x)\geq 0$ (and of course with $\int_{\zone}^{\ztwo}f_*(x)dx = 1$), are obtained by imposing the following conditions,
\begin{eqnarray}
f_*(x=\zeta_-;\zeta_-,\zeta_+)&=&0 \, , \nonumber \\
f_*(x=\zeta_+;\zeta_-,\zeta_+)&=&0 \, .
\label{eq:continuity}
\end{eqnarray}
This yields
\begin{equation}
 \hspace{-1cm}\zeta_{\pm}=\frac{\aone(2+\aone+\atwo)+2\left[1+\atwo\pm\sqrt{(1+\aone)(1+\atwo)
(1+\aone+\atwo)}\right]}{(2+\aone+\atwo)^2} \, ,
\label{eq:zetas}
\end{equation}
which agrees with Eq.~(\ref{eq:zetas}) above, and hence shows that the known results can be recovered using the Coulomb gas approach. In essence this result is related to the expectation that the spectral density vanishes smoothly at the boundaries in the unconstrained case (and this is mathematically expressed by Eqs.~(\ref{eq:continuity})), whereas this is not the case in the constrained cases described below.

Now that we have found the expression for the general spectral density, we will look at four cases of constraints on the spectral density. As will be described below, there are four cases that correspond to the relation between the imposed constraints (sometimes also referred to as walls) $\zone,\ztwo$ and the natural boundaries $\zeta_-,\zeta_+$. In order to simplify the description it would be useful to resort to the Coulomb gas picture, and to envisage the constrained spectral density as the density profile of a collection of charged particles that are put between impenetrable hard walls at $\zone$ and $\ztwo$. In the absence of hard walls, the gas is localized inside $[\zeta_-,\zeta_+]$. This situation is not altered as long as the walls do not perturb the natural boundaries. When $\zone > \zeta_-$ the wall penetrates the lower boundary. The wall is therefore wet by the Coulomb gas, and a new natural upper boundary appears, denoted heres by $U(\zone)$ (obviously shifted to the right with respect to $\zeta_+$). Similarly when $\ztwo < \zeta_+$ the upper wall is wet by the gas and a new lower boundary appears $L(\ztwo)$. The last possibility is when both walls are effective and thus both are wet by the gas.

\subsection{Case 1: $\zone<\zeta_-$ and $\ztwo>\zeta_+$ -- ineffective constraints}
\label{subsec:first}

In this case the imposed constraints $\zone,\ztwo$ lie outside the natural compact support $[\zeta_-,\zeta_+]$ and therefore they do not affect the eigenvalue distribution as the barriers are ineffective. In other words, the walls ($\zone$ and $\ztwo$) do not 'touch' its natural boundaries so nothing happens. This makes the same case as having $\zone=0, \ztwo=1$ or no barriers at all.

Using the explicit values of the natural boundaries, $\zeta_-$ and $\zeta_+$, found in Eq.~\eref{eq:zetas}, and some identities they obey (described in \ref{app:identities}), we can write a simple expression for the unconstrained spectral density
\begin{equation}
f_*(x;\zeta_-,\zeta_+) = \frac{2+\aone+\atwo}{2\pi}\frac{\sqrt{(x-\zeta_-)(\zeta_+-x)}}{x(1-x)} 1_{x \in [\zeta_-,\zeta_+]} \, ,
\label{eq:nobarrier}
\end{equation}
which is already known (see Eq.~(\ref{eq:JMP}) above, and Refs.~\cite{Ledoux,johnstone,Collins}), as it simply recovers the spectral density of the Jacobi ensemble. In essence, apart from re-deriving this expression using the Coulomb gas approach, we show here that in the large $N$ limit conditioning on the eigenvalues being outside the natural support $[\zeta_-,\zeta_+]$ does not affect anything, and in particular the spectral density remains the same.
\Fref{fig:case1} below shows a plot of the unconstrained spectral density when $\aone=2$ and $\atwo=3$.
\begin{figure}[!htb]
\centering%
\includegraphics[scale=.5]{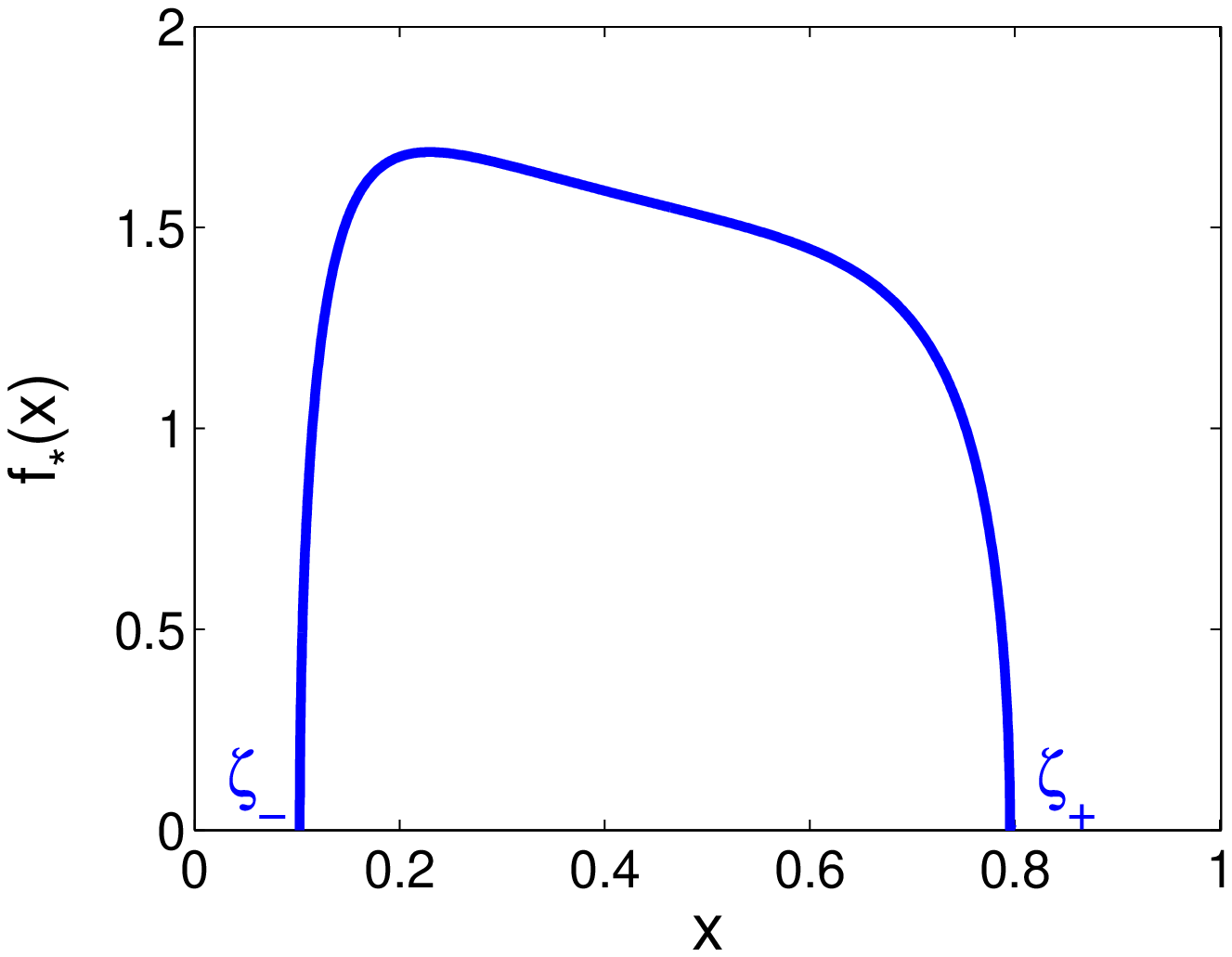}
\caption{Unconstrained spectral density for $\aone=2$ (i.e., $c_1=1/3$) and $\atwo=3$ (i.e., $c_1=1/4$), where $\zeta_-=0.1026...$ and $\zeta_+=0.7953$.}
\label{fig:case1}
\end{figure}

\subsection{Case 2: $\zone \geq \zeta_-$ and $\ztwo \geq U(\zone)$ -- an effective lower barrier}
\begin{figure}[!htb]
\centering%
\includegraphics[scale=.5]{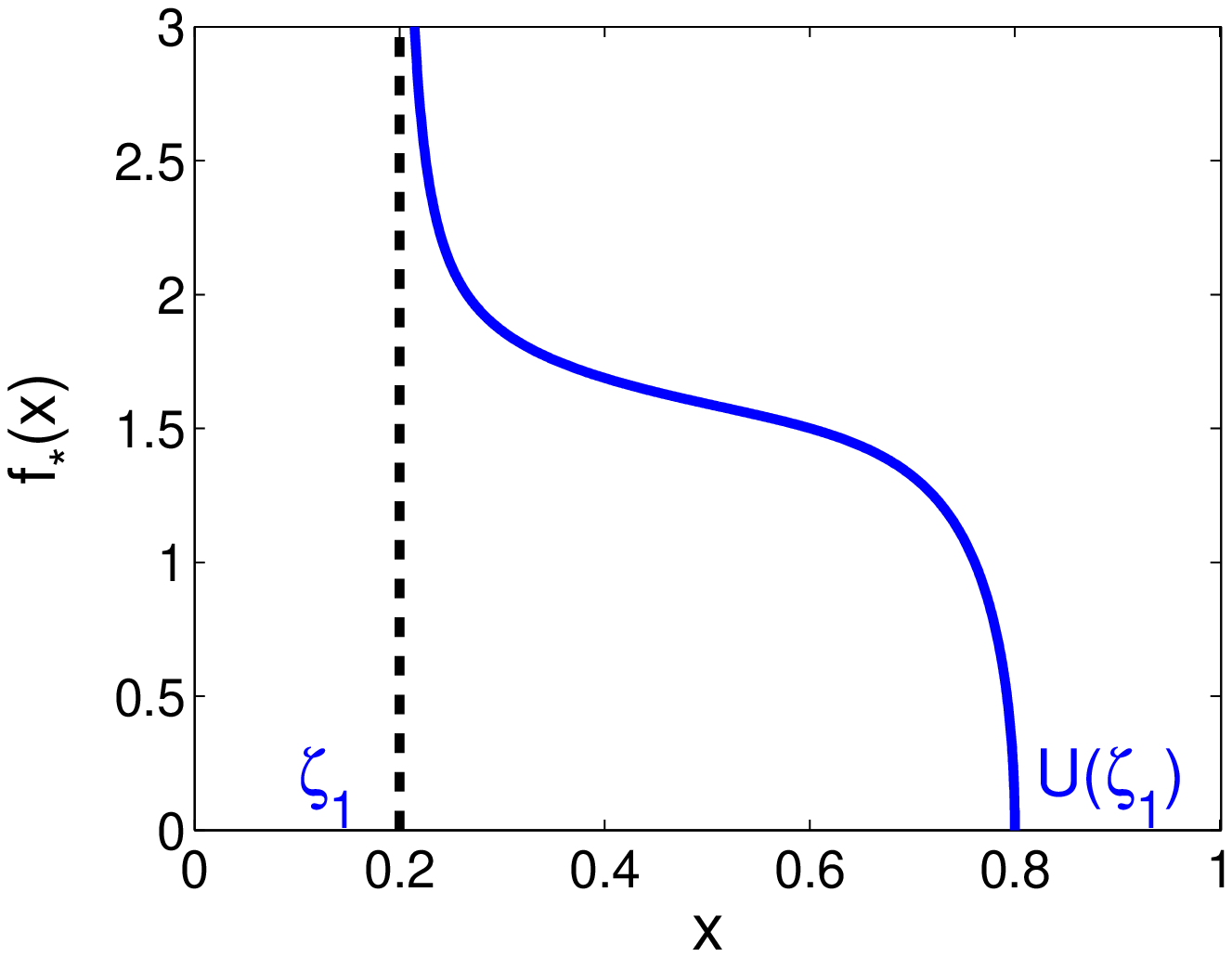}
\caption{The constrained spectral density for $\aone=2$ (i.e., $c_1=1/3$) and $\atwo=3$ (i.e., $c_1=1/4$), while having an effective barrier at $\zone=0.2$ and $U(\zone)=0.8$.}
\label{fig:case2}
\end{figure}

In this case the barrier $\zone$ becomes effective as it exceeds the lower boundary $\zeta_-$. Also, $\ztwo$ is ineffective as it remains $> U(\zone)$. The density is shifted forwards to a new upper boundary $U(\zone)$ leading to the formal density
\begin{equation}
\fl f_*(x;\zone,U(\zone)) = \frac{2+\aone +\atwo}{2\pi} \frac{x-\frac{\aone}{2+\aone+\atwo}\sqrt{\frac{\zone}{U(\zone)}}}{x(1-x)} \frac{\sqrt{U(\zone)-x}}{\sqrt{x -\zone}} 1_{x \in [\zone,U(\zone)]} \, ,
\label{eq:upper}
\end{equation}
obtained from the general expression (\ref{eq:final}) by replacing $\ztwo \rightarrow U(\zone)$.
This density is plotted in \fref{fig:case2} for $\aone=2$ and $\atwo=3$.

Determining the precise value of $U \equiv U(\zone)$ is achieved by requiring $f_*(U;\zone,\ztwo=U)=0$. This leads to the following quartic equation for $w=\sqrt{U}$:
\begin{equation}
w^4 - 2B w^3 + (B^2 + C^2 - 1)w^2 + 2Bw - B^2 = 0 \, ,
 \label{eq:quartic}
\end{equation}
where $B=\gamma_1\sqrt{\zone}$ and $C=\gamma_2\sqrt{1-\zone}$, and where $\gamma_i \equiv \frac{\alpha_i}{2 + \aone + \atwo}$. This equation has four real solutions for all values of $\alpha_i$ and $\zone$ -- one negative and three positive ones. Since we are interested only in positive solutions, and furthermore inside the interval $[0,1]$, we conclude that the right root for $w(\zone)=\sqrt{U(\zone)}$ is the largest solution of $w$, namely
\begin{equation}
U(\zone) = \frac{1}{4} \left[B + \sqrt{z_0} + \sqrt{B^2 - 2C^2 + 2 - z_0 - \frac{2}{\sqrt{z_0}} B (C^2 + 1)} \right]^2 \, ,
\label{eq:U}
\end{equation}
where $z_0$ is the largest root of the cubic resolvent of equation \eref{eq:quartic} \cite{MathWorld}, namely
\begin{equation}
\fl z^3 + (2C^2 - B^2 - 2)z^2 + (1 - C^2)(1 + 2B^2 - C^2)z - B^2(1 + C^2)^2 = 0 \, .
\label{eq:resolvent}
\end{equation}
$z_0$ can be expressed explicitly by
\begin{equation}
z_0 =  - \frac{2C^2 - B^2 - 2}{3} + \left(R + \sqrt{Q^3 + R^2}\right)^{1/3} + \left(R - \sqrt{Q^3 + R^2}\right)^{1/3} \, ,
 \label{eq:z0}
\end{equation}
where $R = \frac{1}{27}\left[ B^6 - 3B^4(1 - C^2) + 3B^2(1 + 16C^2 + C^4) - (1 - C^2)^3 \right]$, \\
and $Q = - \frac{1}{9}(1 - B^2 - C^2)^2$.

An alternative approach to obtain the spectral density that does not require solving a quartic equation and therefore yields much simpler expressions, is to fix $U$, the upper support of the constrained spectral density and then work out the location of the corresponding lower wall $\zone (U)$, using
\begin{equation*}
\zone (U)=U{\left[ \frac{(1-U)\gamma_1+\gamma_2\sqrt{(1-U)\gamma_1^{2}+U\gamma_2^2-U(1-U)}}{\left(1-U \right)\gamma_1^2+U\gamma_2^2} \right]}^2 \, ,
\end{equation*}
for $\frac{1}{2}\left(1 + \gamma_1^2 - \gamma_2^2 + \sqrt{{{\left(1 + \gamma_1^2 - \gamma_2^2 \right)}^2} - 4\gamma_2^2} \right) \le U \le 1$.

Using the results above and making some calculations, we obtain the following expression for the distribution of $\lambda_{min}$:
\begin{equation}
\fl P_N (\lambda_{\min} \ge \zone)=P_N\left(\zone,U(\zone) \right)=\exp\left[-\beta N^2 \Phi_+^{(min)}(\zone - \zeta_-)\right], \quad \zeta_- \leq \zone \leq 1 \, ,
\label{eq:probmin}
\end{equation}
with the right rate function for the positive fluctuations of the minimum eigenvalue, $\Phi_+^{(min)}(x)$, therefore being
\begin{equation}
\Phi_+^{(min)}(x)=\frac{1}{2}\left[S[\zeta_- + x,U(\zeta_- + x)]-S(\zeta_-,\zeta_+)\right], \quad x>0 \, ,
\label{eq:Phi+}
\end{equation}
and where the action $S(\zone,U(\zone)) \equiv S[f_*(x);\zone,U(\zone)]$ is given by:
\begin{eqnarray}
\fl S(\zone,U) &=& -(\aone+\atwo+2) \left(\aone\log\frac{\sqrt{\zone}+\sqrt{U}}{2}
+\atwo\log\frac{\sqrt{1-\zone}+\sqrt{1-U}}{2} \right) \nonumber \\
\fl &+& \frac{{\aone}^2}{2} \log \sqrt{\zone U}+\frac{{\atwo}^2}{2} \log \sqrt{(1-\zone)(1-U)} \nonumber \\
\fl &+& \aone\atwo \log\frac{\sqrt{U(1-\zone)}+\sqrt{\zone(1-U)}}{2}-\log\frac{U-\zone}{4} \, .
\label{eq:finalaction2}
\end{eqnarray}

We have also verified that $U(\zeta_-)=\zeta_+$ as expected, and obtained the following expansion, describing fluctuations of $U$ around $\zeta_+$
\begin{equation}
U(\zeta_- +x) = \zeta_+ +\frac{\textstyle{\left(1-\zeta_+\right){\zeta_+}\over{\left({1-\zeta_-}\right)\zeta_-}}} {4(\zeta_+ - \zeta_-)}x^2 + O\left(x^3\right) \, .
\label{eq:U_fluc}
\end{equation}

\subsection{Case 3: $\ztwo\leq\zeta_+$ and $\zone \leq L(\ztwo)$  -- an effective upper barrier}
\begin{figure}[!htb]
\centering%
\includegraphics[scale=.5]{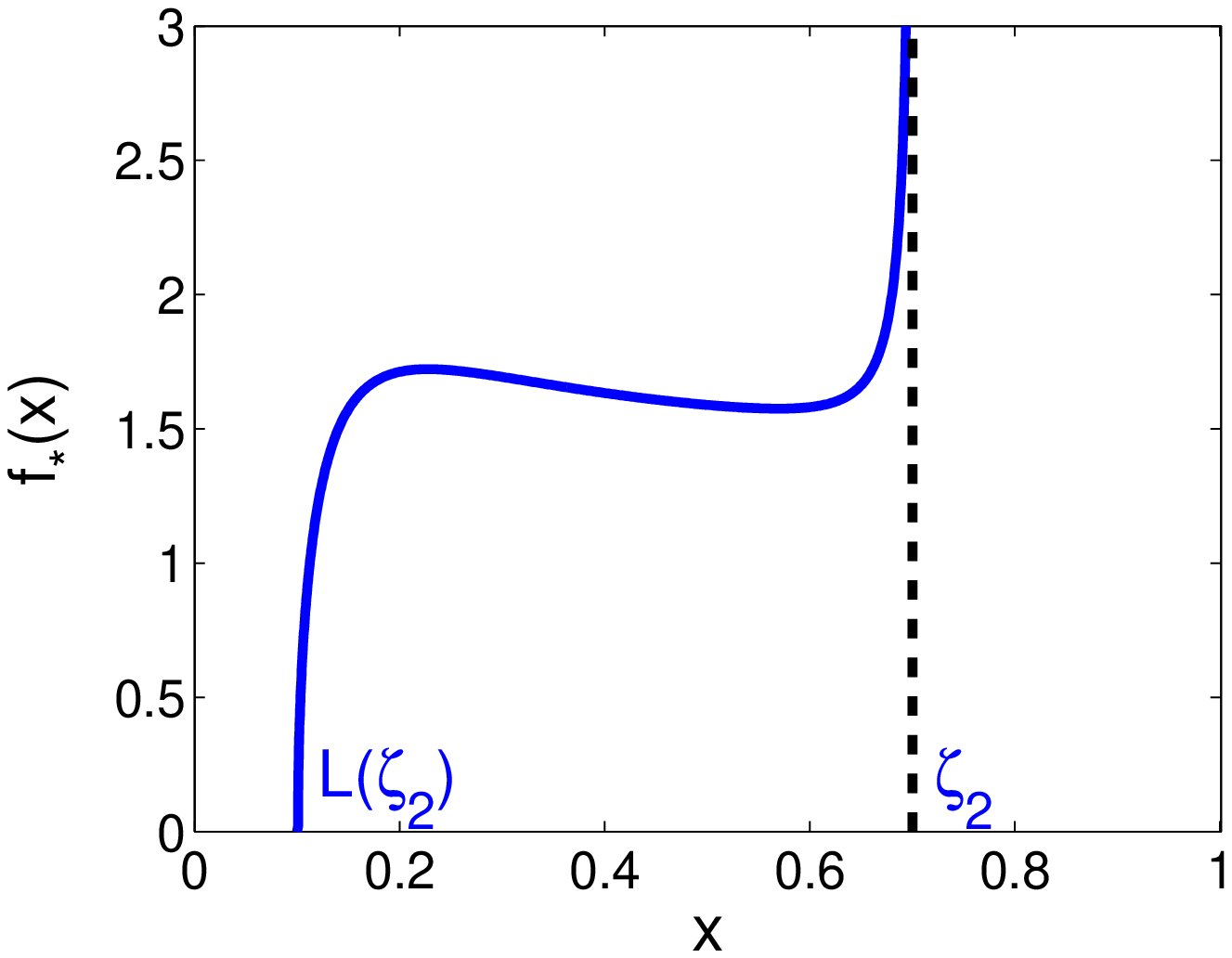}
\caption{The constrained spectral density for $\aone=2$ (i.e., $c_1=1/3$) and $\atwo=3$ (i.e., $c_1=1/4$), while having an effective barrier at $\ztwo=0.7$ and $L(\ztwo) = 0.1009$.}
\label{fig:case3}
\end{figure}

Here, the barrier $\ztwo$ crosses the point $\zeta_+$ from above and so this barrier becomes effective while $\zone$ remains ineffective. In a similar manner as in case 2, the constrained spectral density shifts back to a lower bound value $L(\ztwo)$, which also leads to the same quartic equation as in Eq.~\eref{eq:quartic}, only with $B=\gamma_1\sqrt{\ztwo}$ and $C=\gamma_2\sqrt{1-\ztwo}$. After making the same considerations as before we conclude that the right root for $w(\ztwo)=\sqrt{L(\ztwo)}$ corresponds to the smallest $w$ inside $[0,1]$, and thus,
\begin{equation}
\hspace{-1cm}L(\ztwo) = \frac{1}{4} \left[B + \sqrt{z_0} - \sqrt{B^2 - 2 C^2 + 2 - z_0 - \frac{2}{\sqrt{z_0}} B (C^2 + 1)} \right]^2 \, ,
\label{eq:L}
\end{equation}
where $z_0$ is the largest root of the cubic resolvent Eq.~\eref{eq:resolvent} of the quartic equation \eref{eq:quartic} just as in the previous case. Finally, we can write down the full constrained spectral density as
\begin{equation}
\hspace{-2cm} f_*(x;L(\ztwo),\ztwo) = \frac{2+\aone+\atwo}{2\pi} \frac{\sqrt{x-L\left(\ztwo\right)}}{\sqrt{\ztwo-x}} \frac{\frac{\aone}{2+\aone+\atwo} \sqrt{\frac{\ztwo}{L\left(\ztwo\right)}}-x} {x\left({1-x}\right)} 1_{x \in [L(\ztwo),\ztwo]} \, ,
\label{eq:lower}
\end{equation}
obtained from the general expression (\ref{eq:final}) by replacing $\zone \rightarrow U(\ztwo)$. This density is plotted in \fref{fig:case3} for $\aone=2$ and $\atwo=3$.

An alternative approach to obtain the spectral density that does not require solving a quartic equation and therefore yields much simpler expressions, is to fix $L$, the lower support of the constrained spectral density and then work out the location of the upper wall $\ztwo (L)$, using
\begin{equation*}
\ztwo (L) = L \left[ {\frac{(1 - L)\gamma_1 + \gamma_2\sqrt{\left(1 - L\right)\gamma_1^2 + L\gamma_2^2 - L(1-L)} }{{\left( {1 - L} \right)\gamma _1^2 + L\gamma _2^2}}} \right]^2 \, ,
\end{equation*}
for  $0 \le L \le \frac{1}{2} \left(1 + \gamma _1^2 - \gamma _2^2 - \sqrt{{{\left( {1 + \gamma _1^2 - \gamma _2^2} \right)}^2} - 4\gamma _1^2} \right)$.\\
The probability distribution of $\lambda_{\max}$ can then be expressed as
\begin{equation}
\hspace{-2cm} {P_N}(\lambda_{\max} \le \ztwo)=P_N\left(L(\ztwo),\ztwo \right)=\exp\left[-\beta N^2 \Phi_-^{(max)}(\zeta_+ - \ztwo)\right], \quad 0 \leq \ztwo \leq \zeta_+ \, ,
\label{eq:probmax}
\end{equation}
where the left rate function for the negative fluctuations of the maximum eigenvalue, $\Phi_+^{(max)}(x)$ is
\begin{equation}
\Phi_-^{(max)}(x)=\frac{1}{2}\left[S[L(\zeta_+-x),\zeta_+-x]-S(\zeta_-,\zeta_+)\right], \quad x>0 \, ,
\label{eq:Phi-}
\end{equation}
and the action $S(L,\ztwo)\equiv S[f_*(x);L(\ztwo),\ztwo]$ is
\begin{eqnarray}
\fl S(L,\ztwo) &=& -(\aone+\atwo+2) \left(\aone\log\frac{\sqrt{L}+\sqrt{\ztwo}}{2}
+\atwo\log\frac{\sqrt{1-L}+\sqrt{1-\ztwo}}{2} \right) \nonumber \\
\fl &+& \frac{{\aone}^2}{2} \log \sqrt{L\ztwo}+\frac{{\atwo}^2}{2} \log \sqrt{(1-L)(1-\ztwo)} \nonumber \\
\fl &+&\aone\atwo \log\frac{\sqrt{\ztwo(1-L)}+\sqrt{L(1-\ztwo)}}{2}-\log\frac{\ztwo-L}{4} \, .
\label{eq:finalaction3}
\end{eqnarray}
One can easily verify that $L(\zeta_+)=\zeta_-$, and similar to Eq.~(\ref{eq:U_fluc}) for $U$, we can obtain the following expansion describing fluctuations around $\zeta_-$
\begin{equation}
L(\zeta_+ - x) = \zeta_- - \frac{\textstyle{\left(1-\zeta_-\right){\zeta_-}\over{\left({1-\zeta_+}\right)\zeta_+}}} {4(\zeta_+ - \zeta_-)}x^2 + O\left(x^3\right) \, .
\label{eq:L_fluc}
\end{equation}
Note that by symmetry of the original joint distribution of the eigenvalues (\ref{eq:jpd}) all the expressions in case 3, could have been obtained from those of case 2 by $\aone \leftrightarrow \atwo$, $\zeta \leftrightarrow 1-\zeta$ and also $x \leftrightarrow 1-x$.

\subsection{Case 4: ($\zone>\zeta_-$ and $\ztwo<U(\zone)$) or ($\ztwo<\zeta_+$ and $\zone>L(\ztwo)$) -- two effective barriers}
\begin{figure}[!htb]
\centering%
\includegraphics[scale=.5]{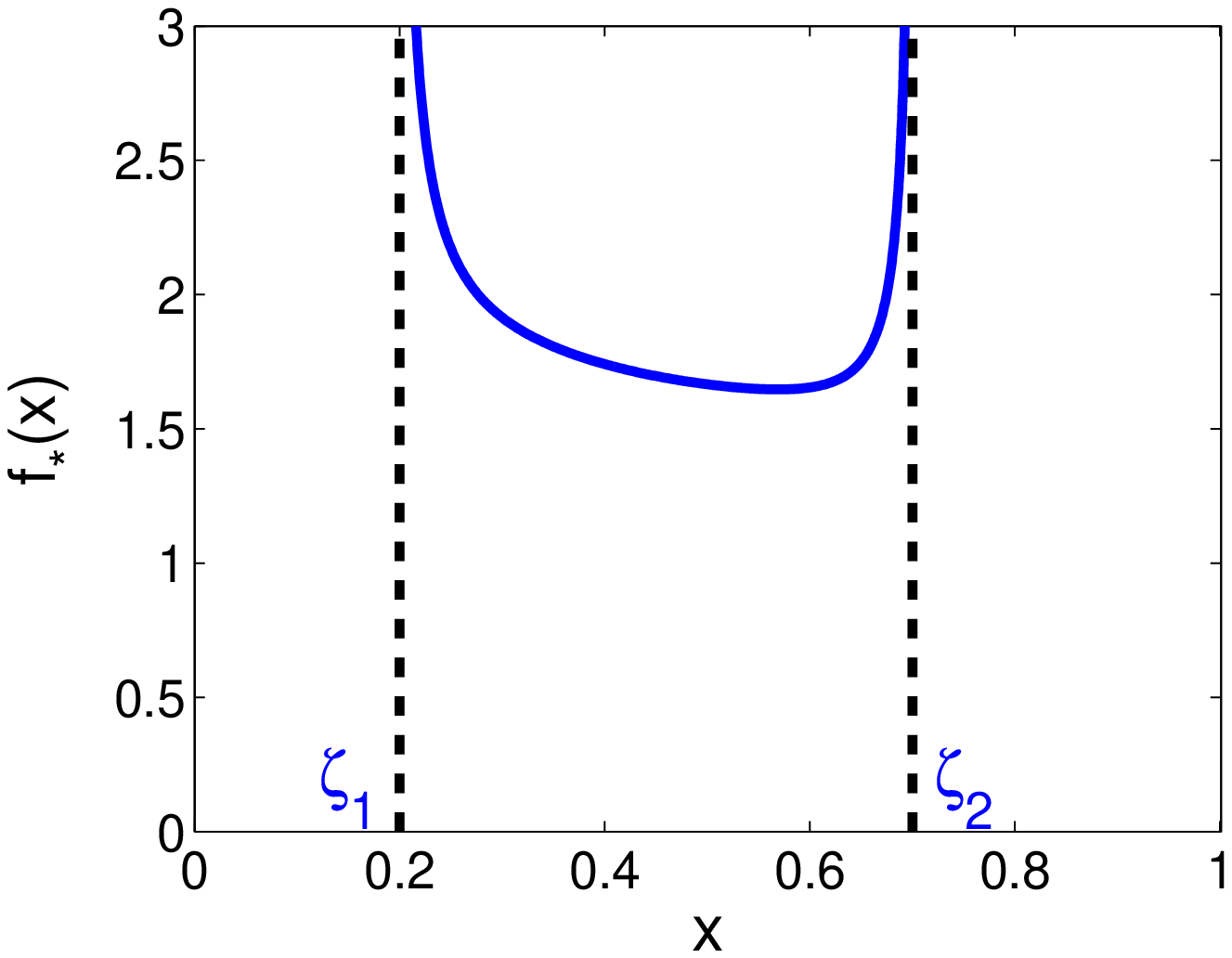}
\caption{The constrained spectral density for $\aone=2$ (i.e., $c_1=1/3$) and $\atwo=3$ (i.e., $c_1=1/4$), with two effective barriers at $\zone=0.2$ and $\ztwo=0.7$}
\label{fig:case4}
\end{figure}

The two barriers become effective since both barriers $\zone$ and $\ztwo$ 'touch' and cross their corresponding natural boundaries. In this case, the spectral density looks just like the general form in \eref{eq:final}, and is repeated here for the sake of completeness:
\begin{eqnarray}
 f_*(x;\zone,\ztwo) &=& \frac{1}{2\pi \sqrt{(x-\zone)(\ztwo-x)}}\Bigg[\aone\left(1-\frac{\sqrt{\zone\ztwo}}{x}\right)\nonumber \\
&+& \atwo\left(1-\frac{\sqrt{(1-\zone)(1-\ztwo)}}{1-x}\right)+2\Bigg] 1_{x \in [\zone,\ztwo]}
\label{eq:case4}
\end{eqnarray}
It is also depicted graphically in Fig.~\ref{fig:case4}.\\
Clearly, the action in this case will be in it's general form as expressed in \eref{eq:finalaction} and we get the partition function \eref{eq:pfzeta}, $Z(\zone,\ztwo) \sim \exp[\beta N^2 S(\zone,\ztwo)/2]$.
For the denominator, $Z(0,1)=Z(\zeta_-,\zeta_+) \sim \exp[\beta N^2 S(\zeta_-,\zeta_+)/2]$, where we have used the fact that the solution for any $\zone<\zeta_-$ and $\ztwo>\zeta_+$ (e.g., when $\zone=0,\ztwo=1$) is the same as the solution for $\zone=\zeta_-$ and $\ztwo=\zeta_+$ (this is exactly case 1 above). Thus, eventually the joint distribution of the extremal eigenvalues $P_N(\zone,\ztwo)$ \eref{eq:jointp} becomes for large $N$
\begin{equation*}
P_N(\zone,\ztwo)=\frac{Z(\zone,\ztwo)}{Z(\zeta_-,\zeta_+)} \sim \exp\left\{-\frac{\beta}{2}N^2\left[S(\zone,\ztwo)-S(\zeta_-,\zeta_+)\right]\right\} \, .
\end{equation*}
where $S(\zeta_-,\zeta_+)$ can be explicitly written as:
\begin{eqnarray}
\hspace{-1cm}  S(\zeta_-,\zeta_+) &=& -(\aone+\atwo+2) \Bigg(\aone\log\frac{\sqrt{(1+\aone)(1+\aone+\atwo)}}{2+\aone+\atwo}\nonumber\\
\hspace{-1cm}&+&\atwo\log\frac{\sqrt{(1+\atwo)(1+\aone+\atwo)}}{2+\aone+\atwo} \Bigg)+ \frac{{\aone}^2}{2} \log \frac{\aone}{2+\aone+\atwo}\nonumber\\
\hspace{-1cm}&+&\frac{{\atwo}^2}{2} \log \frac{\atwo}{2+\aone+\atwo}+\aone\atwo \log\frac{\sqrt{(1+\aone)(1+\atwo)}}{2+\aone+\atwo}\nonumber\\
\hspace{-1cm}&-&\log\frac{\sqrt{(1+\aone)(1+\atwo)(1+\aone+\atwo)}}{(2+\aone+\atwo)^2} \, .
\label{eq:finalaction4}
\end{eqnarray}

\subsection{Summary of the joint distribution $P_N(\lambda_{\min} \ge \zone, \lambda_{\max} \le \ztwo)$}
It is useful to summarize the results derived above for the joint distribution $P_N(\lambda_{\min} \ge \zone, \lambda_{\max} \le \ztwo)$ in one formula in order to highlight the general structure, and to facilitate applications:
\begin{equation*}
\begin{array}{l}
 \fl {P_N}\left(\lambda_{\min} \ge \zone,{\lambda_{\max}} \le \ztwo \right) \sim  \\
 \quad \sim \left\{ \begin{array}{l}
 1\quad \quad \quad \quad \quad \quad \quad \quad \quad \quad \quad \quad \quad \quad \zone \le \zeta_- \,\, {\rm{and}} \,\, \ztwo \ge \zeta_+ \\
 {{\mathop{\rm e}\nolimits} ^{ - \frac{\beta }{2}{N^2}\left\{ {S\left[ {{f_*}\left( x \right);L\left( {\ztwo} \right),\ztwo} \right] - S\left( {\zeta_-,\zeta_+} \right)} \right\}}}\quad \quad \ztwo < \zeta_+ \,\, {\rm{and}} \,\, \zone \le L\left( {\ztwo} \right) \\
 {{\mathop{\rm e}\nolimits} ^{ - \frac{\beta }{2}{N^2}\left\{ {S\left[ {{f_*}\left( x \right);\zone,U\left( {\zone} \right)} \right] - S\left( {\zeta_-,\zeta_+} \right)} \right\}}}\quad \quad \zone > \zeta_- \,\, {\rm{and}} \,\, \ztwo \ge U\left( {\zone} \right) \\
 {{\mathop{\rm e}\nolimits} ^{ - \frac{\beta }{2}{N^2}\left\{ {S\left[ {{f_*}\left( x \right);\zone,\ztwo} \right] - S\left( {\zeta_-,\zeta_+} \right)} \right\}}}\quad \quad \quad {\rm Otherwise} \\
 \end{array} \right. \\
 \end{array} \, .
\end{equation*}
where the $\sim$ symbol stands for logarithmic equivalence for large $N$, or to be more precise, to order ${\vO}(N^2)$ in the exponential, with possible ${\vO}(N)$ corrections. The various expressions for the action $S$ are given by Eqs.~(\ref{eq:finalaction}), (\ref{eq:finalaction2}), (\ref{eq:finalaction3}) and (\ref{eq:finalaction4}), $U(\zone)$ given by Eq.~(\ref{eq:U}) and $L(\ztwo)$ given by Eq.~(\ref{eq:L}).

\section{The other rate functions - ${\vO}(N)$ fluctuations on the other side}
\label{sec:othersides}
The saddle-point approximation that we have used to evaluate $Z(\zone,\ztwo)$ is only able to capture the large fluctuations of $\lambda_{\rm min}$ to the right of its mean value $\langle \lambda_{\rm min} \rangle =\zeta_-$ (and similarly the large fluctuations of $\lambda_{\rm max}$ to the right of its mean value $\langle \lambda_{\rm max} \rangle =\zeta_+$). Fortunately, the authors of \cite{wishartgaussian} came up with a beautiful physical argument to overcome this shortcoming and to estimate in their case the large deviations of the largest eigenvalue of the Wishart ensemble. In the following we will apply this method to the Jacobi ensemble, and in particular to the fluctuation to the right of $\lambda_{\rm max}$ as well as to the fluctuation to the left of $\lambda_{\rm min}$.

\subsection{$\Phi_+^{(\max)}(x)$: the right rate function of $\lambda_{\rm max}$}

We will start with the derivation of the right rate function for the large positive fluctuations of $\lambda_{\rm max}$, namely $\Phi_+^{(\max)}(x)$, defined using the following tail distribution
\begin{equation*}
P_N^{(\max)}(\zeta) \sim e^{-\beta N\Phi_{+}^{(\max)}(\zeta - \zeta_+)}, \quad \quad \zeta \in [\zeta_+,1] \, .
\end{equation*}
Note that the fluctuations in this case are of order ${\vO}(N)$ while the fluctuations to the left of $\zeta_+$ are of order ${\vO}(N^2)$ - see Eq.~(\ref{eq:maxm}). In a nutshell, the reason for this difference between the right and left fluctuations is that for $\lambda_{\rm max}$ to be smaller than its mean value it requires ALL the $N$ eigenvalues to be smaller than that mean value, while for $\lambda_{\rm max}$ to be larger than its mean value has no such mandatory effect on the other eigenvalues. Equivalently, using the Coulomb-gas language, pushing $\lambda_{\rm max}$ to the left compresses all the particles, while pulling $\lambda_{\rm max}$ to the right does not necessarily moves the others. The explains a difference in factor $N$ between the energetic contributions of the two states, as explained below.

The starting point of the method \cite{wishartgaussian} to compute $\Phi_+^{(\max)}(x)$ is the energy expression \eref{eq:freeenergy}. The Coulomb gas physics suggests that when the rightmost charge is moved to the right from $\zeta_+$, i.e. $\lambda_{\rm max}- \zeta_+ \sim {\vO}(1)$, the sea of charges is {\it a priori} not subject to forces capable of macroscopic rearrangements. Following this physical picture, the right rate function is determined by the energy cost in pulling the rightmost charge in the external potential of the Coulomb gas and the interaction of the charge with the {\it unperturbed} sea. This energy cost for $\lambda_{\rm  max}=\zeta \gg \zeta_+$ can be estimated for large $N$ using \Eref{eq:freeenergy}
\begin{equation}
\Delta E(\zeta)= -\aone \ln(\zeta) -\atwo\ln(1-\zeta) -2\int \ln|\zeta-\lambda|\, \rho_N(\lambda)\,d\lambda\,,
\label{eq:ec1}
\end{equation}
where $\rho_N(\lambda)$ is the average density of charges and the integral describes the Coulomb interaction of the rightmost charge with the sea. For large $N$, we use the expression for the spectral density \eref{eq:final} and the energy cost finally takes the form
\begin{equation}
\Delta E(\zeta) = -\aone\ln \zeta - \atwo\ln(1-\zeta) -2\int\limits_{\zeta_-}^{\zeta_+} {\ln \left(\zeta - \zeta'\right) f_*(\zeta')d\zeta'}  + C \, .
\label{eq:ec2}
\end{equation}
This energy cost expression is valid up to an additive constant $C$, which will be chosen here such that $\Delta E (\zeta=\zeta_+)=0$ since our reference configuration is the one where $\lambda_{\rm max}= \zeta_+$. We can now estimate the probability of such a configuration by $P_N(\zeta)\propto\exp\left[-\beta\Delta E (\zeta)/2\right]$.\\
Using the integrals \eref{eq:J1} \eref{eq:J2} detailed in \ref{app:case1}, we can derive the explicit expression for the right rate function
$\Phi_{+}^{(\max)}(x)=\frac{1}{2} \Delta E(\zeta_+ + x)$
\begin{eqnarray}
 \hspace{-2.3cm}\Phi_{+}^{(\max)}(x) &=& \frac{\aone}{2} \ln\left(1+\frac{x}{\zeta_+}\right) + \frac{\atwo}{2} \ln \left(1 - \frac{x}{1-\zeta_+} \right) \nonumber\\
 \hspace{-2.3cm}&-& (2+\aone+\atwo)\ln \left(\sqrt{1 +\frac{x}{\Delta_-}} + \sqrt{\frac{x}{\Delta_-}} \right)- \aone\ln \left({\sqrt{1+\frac{x}{\Delta_-}} - \sqrt\frac{\zeta_-}{\zeta_+} \sqrt{\frac{x}{\Delta_-}}}\right) \nonumber \\
 \hspace{-2.3cm}&-&\atwo\ln\left({\sqrt{1+\frac{x}{\Delta_-}} - \sqrt{\frac{1-\zeta_-}{1-\zeta_+}} \sqrt{\frac{x}{\Delta_-}}}\right) \, ,
\label{eq:rldv-Jacobi}
\end{eqnarray}
where $x \in \left[0,(1-\zeta_+)\right]$, and as mentioned above the additive constant was chosen to have $\Phi_{+}^{(\max)}(0)=0$. Also, we use the notation $\Delta_-=\zeta_+ - \zeta_-$ (a simplified expression for $\Delta_-$ is available in \ref{app:identities}).\\
We now check that the smallest large fluctuations predicted by our results match the largest typical fluctuations given by the Tracy--Widom distribution \cite{johnstone} (as long as $c_1,c_2<1$, i.e. for rectangular matrices). Expanding the rate functions $\Phi_{\pm}^{(\max)}(x)$ around $x=0$ gives:
\begin{equation*}
\Phi_{+}^{(\max)}(x) \stackrel{x\to0}\sim \frac{2}{3} \frac{\left[(1+\aone)(1+\atwo)(1+\aone+\atwo) \right]^{1/4}}{\zeta_+(1-\zeta_+)} \, x^{3/2}+ \, \Or (x^{5/2})
\end{equation*}
and from Eqs.~(\ref{eq:Phi-})-(\ref{eq:finalaction3})
\begin{eqnarray*}
\Phi_{-}^{(\max)}(x) \stackrel{x\to0}{\sim} \frac{1}{24} \frac{\sqrt{(1+\aone)(1+\atwo)(1+\aone+\atwo)}}{\left[\zeta_+ (1-\zeta_+)\right]^2} \, x^3+ \, \Or (x^4) \, .
\end{eqnarray*}
This yields the following expression of $P^{(\max)}_{N}(\zeta)$ for the smallest large fluctuations of $\lambda_{\rm max}$:
\begin{equation*}
P^{(\max)}_{N}(\zeta) \sim \cases{\exp\left(- {\textstyle{{2\beta} \over 3}} \chi^{3/2}(\zeta)\right), \quad \zeta\in [\zeta_{+},1] \\
\exp\left(- {\textstyle{{\beta} \over 24}} |\chi(\zeta)|^3\right), \quad \zeta \in [0,\zeta_{+}]}
\end{equation*}
with
\begin{equation*}
\chi(\zeta) \equiv \frac{[(1+\aone)(1+\atwo)(1+\aone+\atwo)]^{1/6}}{\left[\zeta_+ (1-\zeta_+)\right]^{2/3}} \frac{(\zeta - \zeta_+)}{N^{1/3}} \, ,
\end{equation*}
where the symbol $\sim$ in the last expression stand for log-equivalence for large $N$.
This result agrees with Tracy--Widom distribution for large $|\chi|$ \cite{johnstone,johnstone2,TW} - see also \ref{app:TW}.\\
Last, for larger atypical fluctuations, we have the following asymptotic behaviour of the rate function
\begin{eqnarray*}
\hspace{-2cm} \Phi_{+}^{(\max)}(x) && \stackrel{x\to 1-\zeta_+} \sim - \frac{\atwo}{2} \ln(1-\zeta_+ - x) + \left[\atwo \ln{\atwo} - \frac{1+\atwo}{2}\ln(1+\atwo) \right. \nonumber \\
\hspace{-2cm} &+& \left. \frac{1+\aone}{2}\ln(1+\aone) - \frac{1+\aone+\atwo}{2}\ln(1+\aone+\atwo) \right] + O(1-\zeta_+ - x)\, .
\end{eqnarray*}
\subsection{$\Phi_-^{(\min)}(x)$: the left rate function of $\lambda_{\rm min}$}
We now derive the analogous formula for the large fluctuations of the smallest eigenvalue $\lambda_{\rm min}$, namely $\Phi_{-}^{(\min)}(x)$, which is defined using the following tail distribution
\begin{equation*}
P_N^{(\min)}(\zeta) \sim e^{-\beta N\Phi_-^{(\min)}(\zeta_- - \zeta)}, \quad \quad \zeta \in [0,\zeta_-] \, .
\end{equation*}
Using the same ideas as above, similar steps and integrals (more specifically \eref{eq:J3} \eref{eq:J4} from \ref{app:case1}) we arrive at the following final expression for the rate function
\begin{eqnarray}
\hspace{-2.3cm} \Phi_{-}^{(\min)}(x) &=& \frac{\aone}{2} \ln \left(1-\frac{x}{\zeta_-} \right)+\frac{\atwo}{2}\ln\left(1 + \frac{x}{1-\zeta_-} \right)\nonumber\\
&-&(2+\aone+\atwo)\ln\left(\sqrt{1+\frac{x}{\Delta_-}} + \sqrt{\frac{x}{\Delta_-}}\right) - \aone\ln \left({\sqrt{1+\frac{x}{\Delta_-}} -\sqrt\frac{\zeta_+}{\zeta_-} \sqrt{\frac{x}{\Delta_-}}}\right)\nonumber \\
\hspace{-2.3cm}& -&\atwo\ln\left({\sqrt{1 + \frac{x}{\Delta_-}}-\sqrt{\frac{1-\zeta_+}{1-\zeta_-}} \sqrt{\frac{x}{\Delta_-}} }\right) \, .
\label{eq:lldv-Jacobi}
\end{eqnarray}
Expanding the rate functions $\Phi_{\pm}^{(\min)}(x)$ around $x=0$ gives:
\begin{equation*}
\Phi_{-}^{(\min)}(x) \stackrel{x\to0}\sim \frac{2}{3} \frac{\left[(1+\aone)(1+\atwo)(1+\aone+\atwo) \right]^{1/4}}{\zeta_-(1-\zeta_-)} \, x^{3/2} + \, \Or(x^{5/2})  \, ,
\end{equation*}
and from Eqs.~(\ref{eq:Phi+})-(\ref{eq:finalaction2})
\begin{eqnarray*}
\Phi_{+}^{(\min)}(x) \stackrel{x\to0} \sim \frac{1}{24} \frac{\sqrt{(1+\aone)(1+\atwo)(1+\aone+\atwo)}}{\left[\zeta_- (1-\zeta_-)\right]^2} \, x^3 +\,\Or(x^{4}) \, ,
\end{eqnarray*}
which yield the following expression of $P^{(\min)}_{N}(\zeta)$ for the smallest large fluctuations of $\lambda_{\rm min}$:
\begin{equation*}
P^{(\min)}_{N}(\zeta)\sim\cases{\exp\left(- {\textstyle{{2\beta } \over 3}} \chi^{3/2}(\zeta)\right),\quad \zeta \in [0,\zeta_{-}] \\
\exp\left(- {\textstyle{{\beta } \over 24}} |\chi(\zeta)|^3\right),\quad \zeta \in [\zeta_{-},1]}
\end{equation*}
with
\begin{equation*}
\chi(\zeta) \equiv - \frac{[(1+\aone)(1+\atwo)(1+\aone+\atwo)]^{1/6}}{\left[\zeta_- (1-\zeta_-)\right]^{2/3}} \frac{(\zeta_- - \zeta)}{N^{1/3}}\, .
\end{equation*}
This result again agrees with the Tracy--Widom distribution for large $|\chi|$ \cite{johnstone,johnstone2,TW} - see also \ref{app:TW}.\\
Last, for larger atypical fluctuations, we have the following asymptotic behaviour of the rate function
\begin{eqnarray*}
\fl \Phi_{-}^{(\min)}(x) \stackrel{x\to \zeta_-} \sim &-&\frac{\aone}{2} \ln(\zeta_- - x) + \left[\aone \ln{\aone} - \frac{1+\aone}{2}\ln(1+\aone) \right. \nonumber \\
\fl &+& \left. \frac{1+\atwo}{2}\ln(1+\atwo) - \frac{1+\aone+\atwo}{2}\ln(1+\aone+\atwo) \right] + O(\zeta_- - x)\, .
\end{eqnarray*}

\section{Numerical Results}
\label{sec:num}
Formulas \eref{eq:final} and \eref{eq:zetas} have been numerically checked using MATLAB and the agreement with the analytical results is already excellent. In Fig. 5, we plot the histograms of normalized eigenvalues $\lambda$ for different values of $N$, up to $N=100$, with an initial sample of $10^6$ Jacobi matrices $(\beta=1, c_1=1/3, c_2=1/4)$. On top of each histogram, we plot the corresponding theoretical distribution given by Eq.~(\ref{eq:nobarrier}). An important reason for presenting this figure is to give an idea about the convergence rate of the spectral distribution to is large-$N$ limit.\\
\begin{figure}[ht]
\centering
\subfigure[$(N,M_1,M_2)=(5,15,20)$]{
\includegraphics[scale=.5]{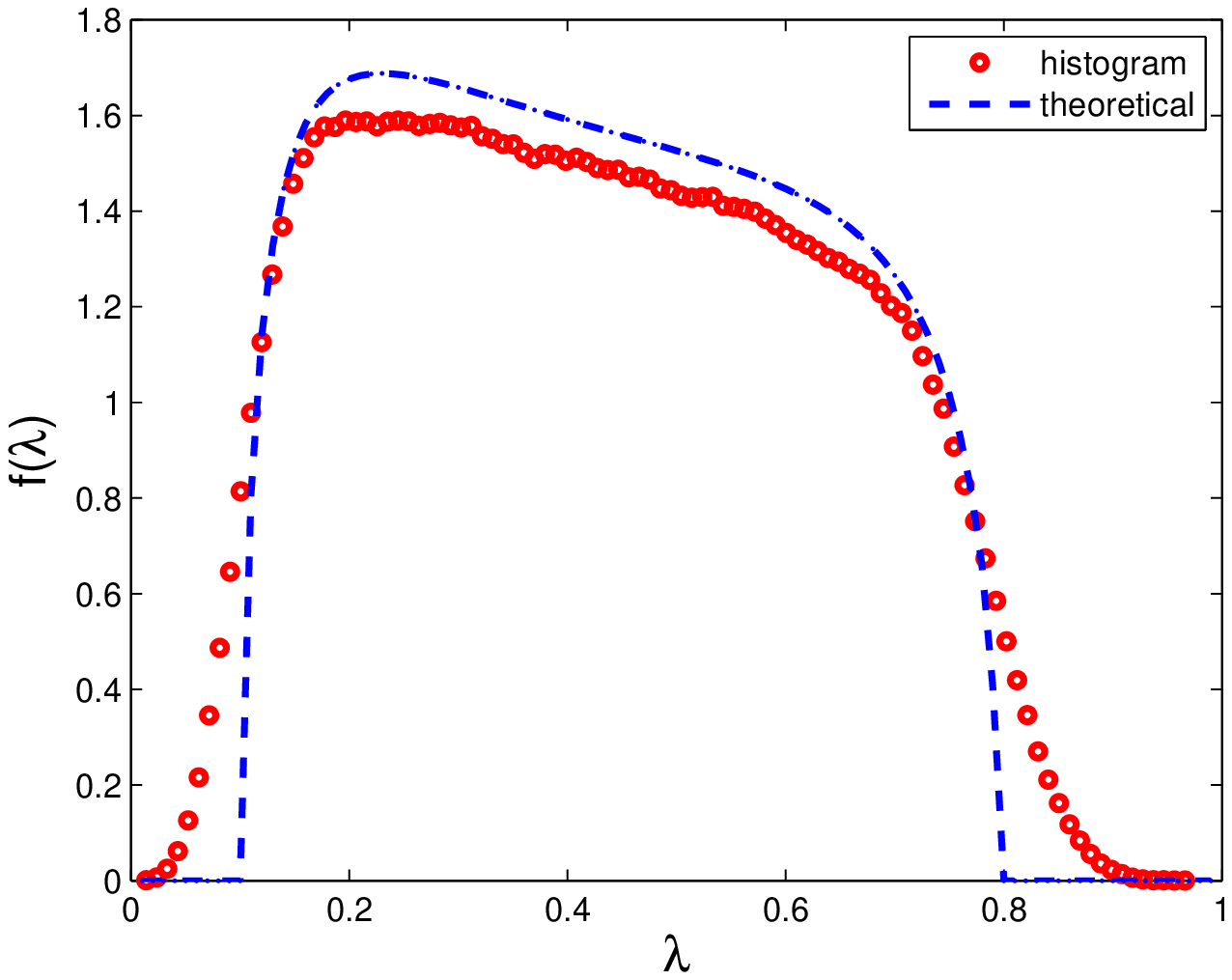}
\label{fig:spec5}
}
\subfigure[$(N,M_1,M_2)=(10,30,40)$]{
\includegraphics[scale=.5]{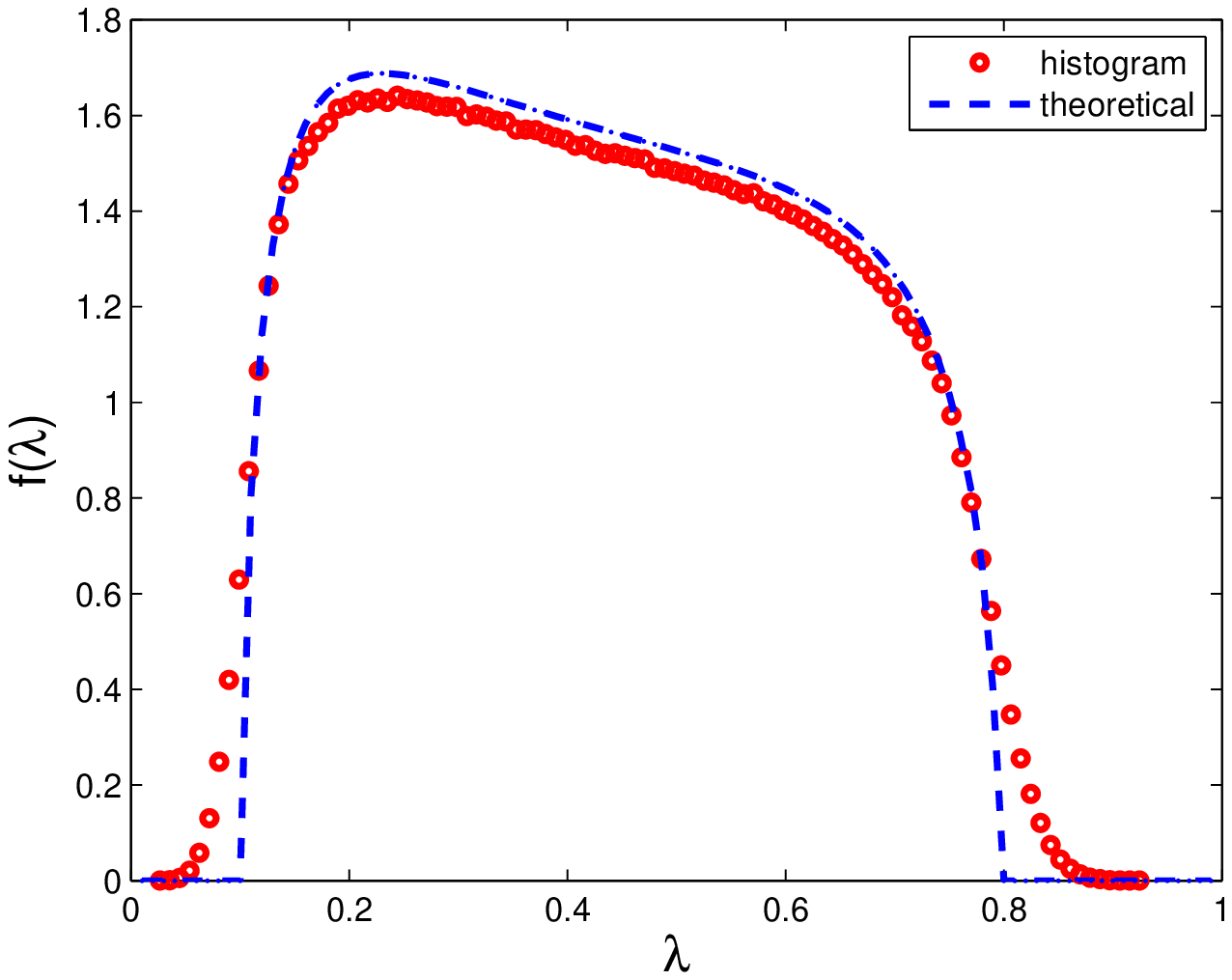}
\label{fig:spec10}
}
\subfigure[$(N,M_1,M_2)=(30,90,120)$]{
\includegraphics[scale=.5]{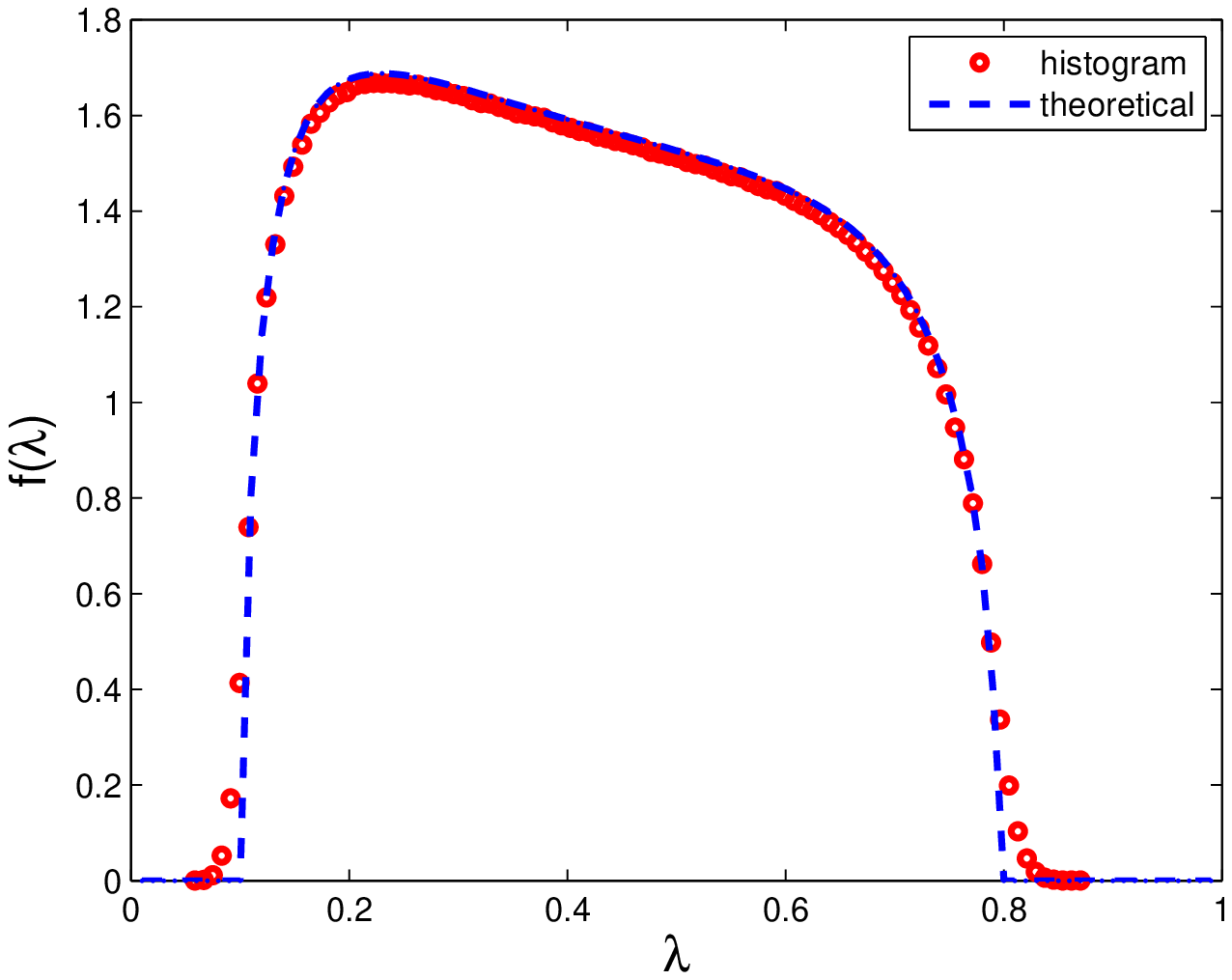}
\label{fig:spec30}
}
\subfigure[$(N,M_1,M_2)=(100,300,400)$]{
\includegraphics[scale=.5]{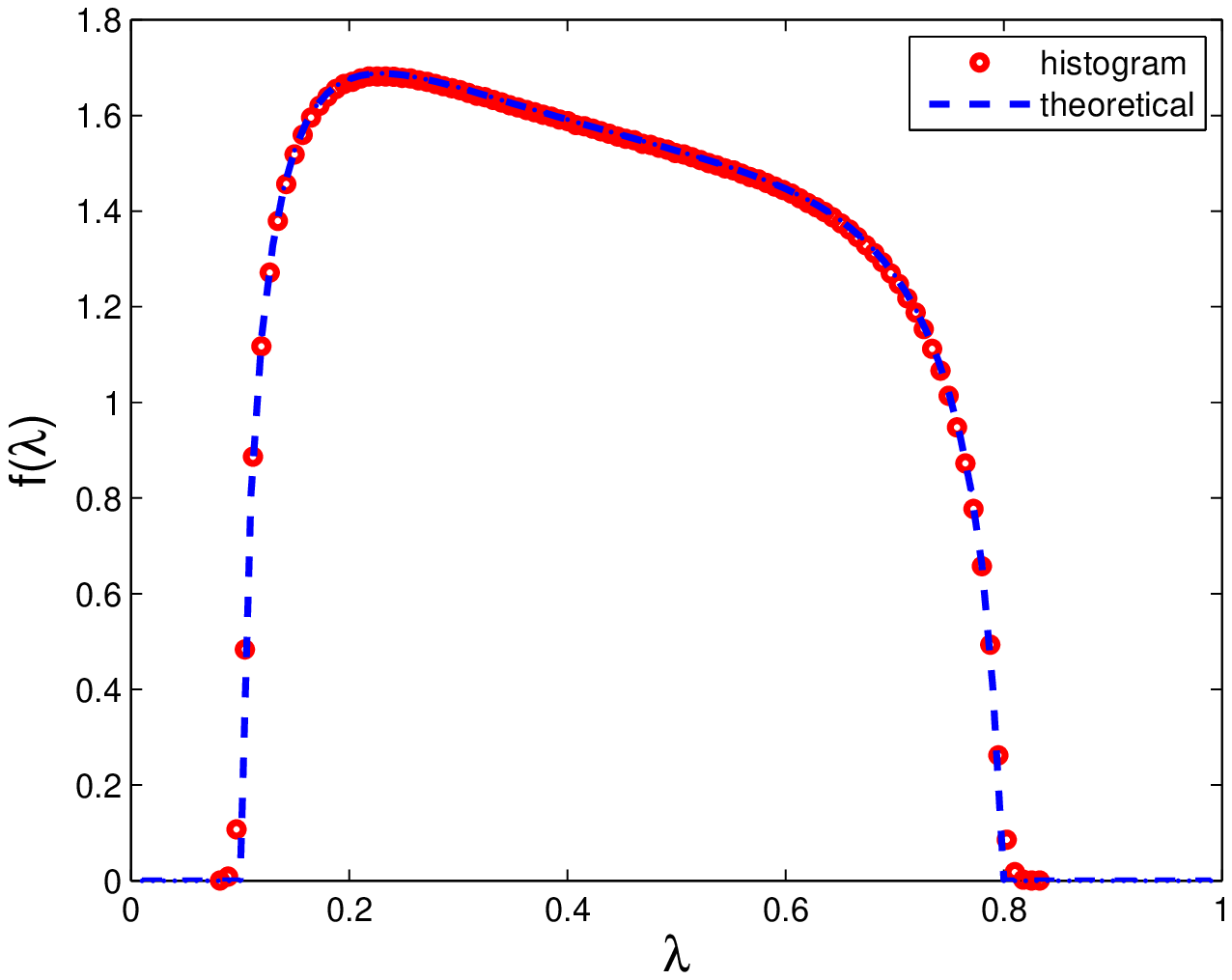}
\label{fig:spec100}
}
\label{fig:spectrals}
\caption{The red circles in the subfigures represent the normalized histograms of eigenvalues of Jacobi ensembles that have been left to run for $10^6$
repetitions. From top left to bottom right, the figure shows how the theoretical limiting spectral density (blue dashed line) converges gradually with the
corresponding simulation results (red circles) as we increase our $N$ values, with $\aone=2$ ($c_1 = 1/3$) and $\atwo=3$ ($c_2 =1/4$). At large $N=100$ in subfigure\subref{fig:spec100}, the graphs can be seen to align precisely.}
\end{figure}
Next, we performed numerical simulations to confirm the correctness of the formulas for the constrained spectral densities. For this purpose we have used the so called tridiagonal models for $\beta$-Jacobi ensembles \cite{Suttonthesis,Lippert,Dumitriu}, described in the following. The underlying observation (proven in \cite{Suttonthesis}) is that the tridiagonal positive semidefinite matrix $J=Z Z^T$ is $\beta$-Jacobi distributed (i.e., has exactly the same joint probability distribution of eigenvalues as (\ref{eq:jpd})), where $Z$ is the following upper bidiagonal $N \times N$ matrix
\begin{equation}
Z \sim \left( {\begin{array}{*{20}{c}}
   {c_N} & { - {s_N}{{c'}_{N - 1}}} & {} & {} & {}  \\
   {} & {{c_{N - 1}}{{s'}_{N - 1}}} & {-s_{N - 1}{c'}_{N - 2}} & {} & {}  \\
   {} & {} & {{c_{N - 2}}{{s'}_{N - 2}}} &  \ddots  & {}  \\
   {} & {} & {} &  \ddots  & { - {s_2}{{c'}_1}}  \\
   {} & {} & {} & {} & {{c_1}{{s'}_1}}  \\
\end{array}} \right) \, ,
\label{eq:Z}
\end{equation}
where all $c_i$'s and $c'_i$'s are independent random variables distributed as follows:
\begin{eqnarray}
\hspace{-2cm} {c_i} \sim \sqrt{Beta\left( {{\textstyle{\beta  \over 2}} (N\aone + i),{\textstyle{\beta \over 2}}(N\atwo + i)} \right)}, \qquad && s_i = \sqrt{1-c_i^2} \quad 1 \le i \le N \, , \\
\hspace{-2cm} {c'_i} \sim \sqrt{Beta\left( {\textstyle{\beta  \over 2}i,{\textstyle{\beta \over 2}}(N\aone + N\atwo + 1 + i)} \right)} , \qquad && {s'}_i = \sqrt{1-{c'}_i^2} \quad 1 \le i \le N-1 \, ,
\label{eq:ci}
\end{eqnarray}
where $Beta(s,t)$ stands for the beta distribution, whose density is proportional to $x^{s-1}(1-x)^{t-1}$ on $[0,1]$.\\
Figure~\ref{fig:Nspectrals} shows a comparison of various spectral densities simulated using the tridiagonal trick and our theoretical predictions based on the Coulomb gas picture. We used here complex ($\beta=2$) Jacobi matrices ($N=30$) with $\aone=0.6, \atwo=0.8$. Fig.~\ref{fig:Ncase1} show the unconstrained spectral density with a good agreement with the numerical simulations (we used here $10^5$ samples of matrices). The other three insets describe the case with an effective lower barrier $\zone$ (Fig.~\ref{fig:Ncase2}), an effective upper barrier $\ztwo$ (Fig.~\ref{fig:Ncase3}), and two effective barriers (Fig.~\ref{fig:Ncase4}).\\
\begin{figure}[ht]
\centering
\subfigure[Unconstrained]{
\includegraphics[scale=.4]{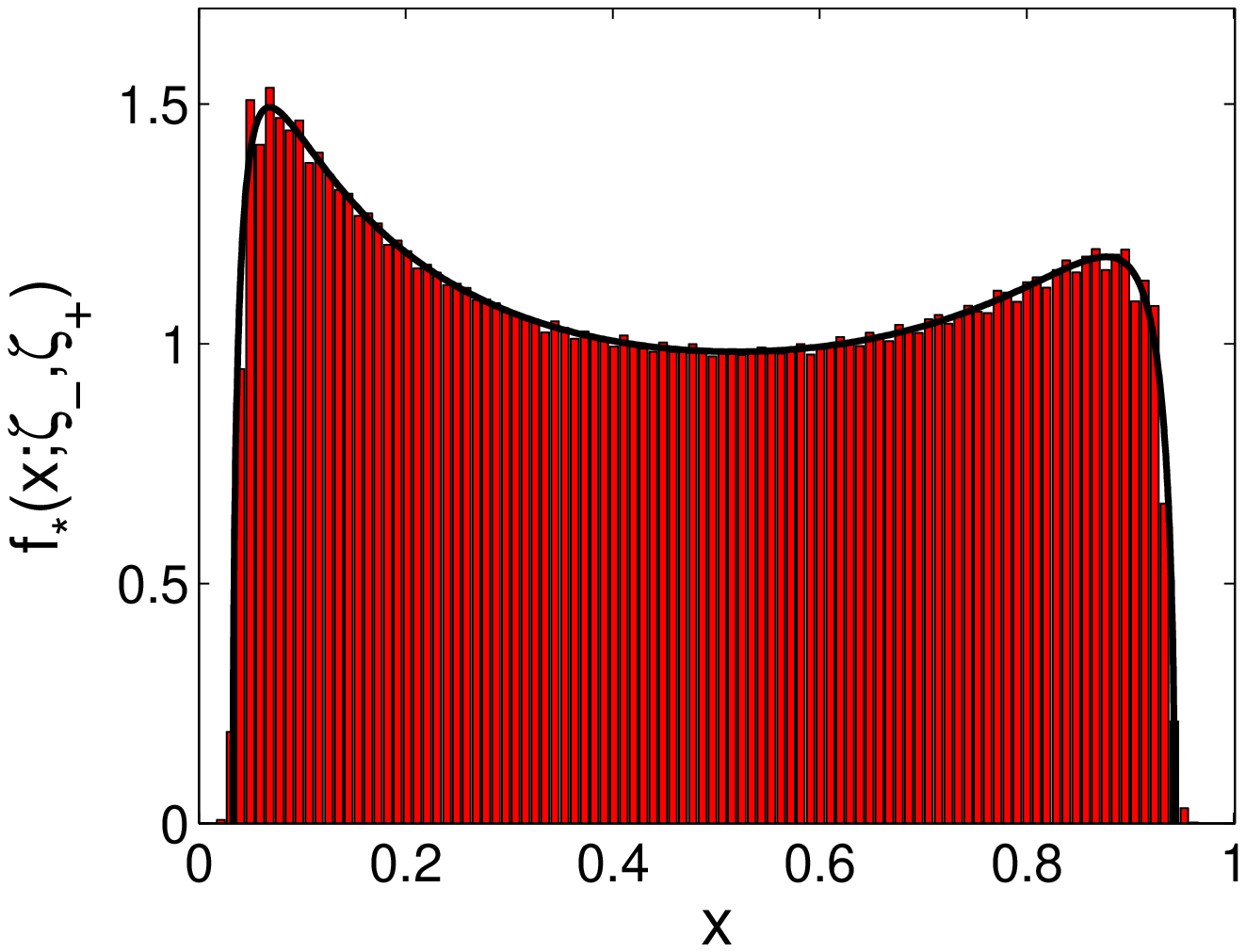}
\label{fig:Ncase1}
}
\subfigure[$\zone=0.07 \, , \ztwo=U(0.07) \simeq 0.9431$]{
\includegraphics[scale=.4]{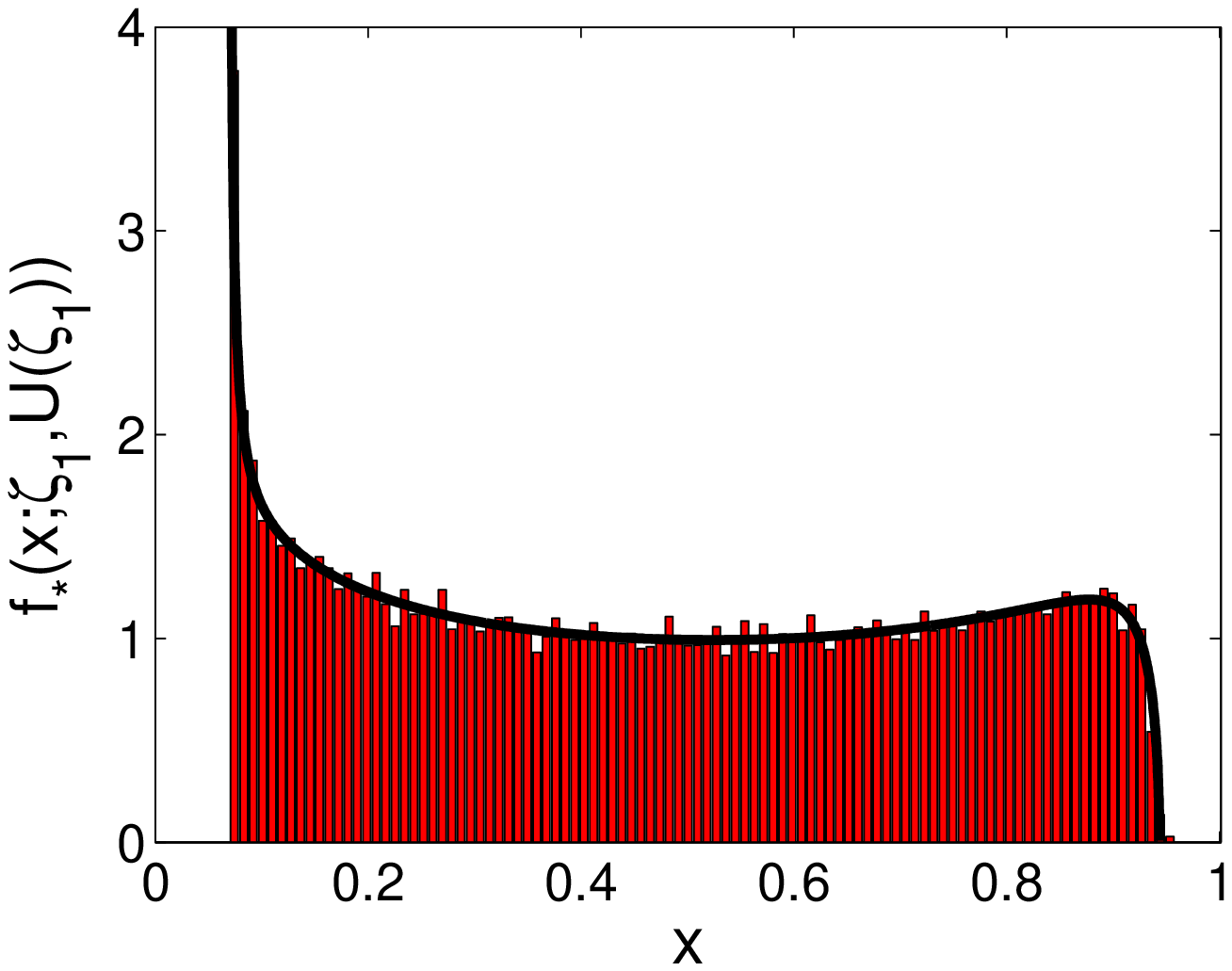}
\label{fig:Ncase2}
}
\subfigure[$\ztwo=0.90 \, , \ztwo=L(0.90) \simeq 0.0328$]{
\includegraphics[scale=.4]{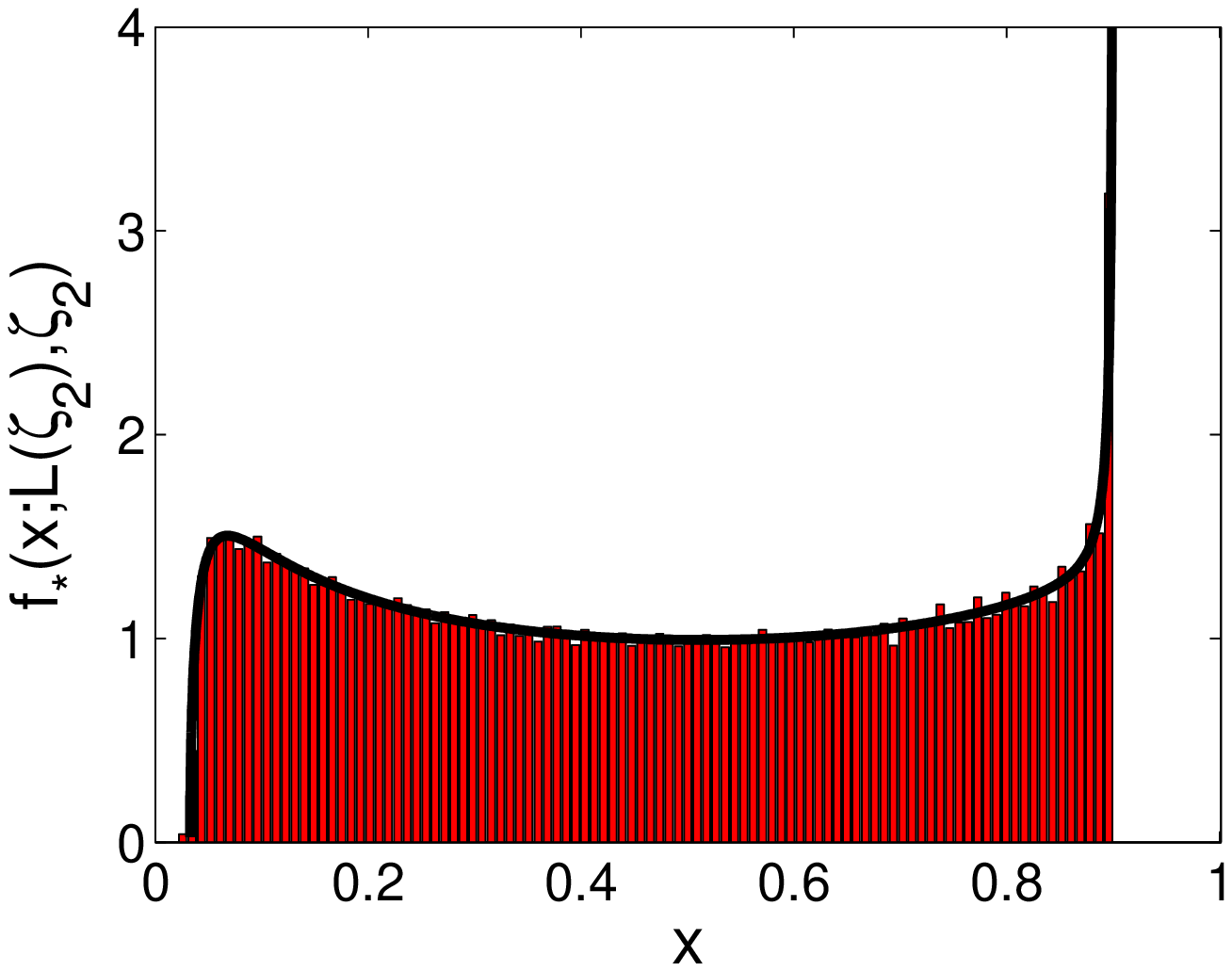}
\label{fig:Ncase3}
}
\subfigure[$\zone=0.05 \, , \ztwo=0.90$]{
\includegraphics[scale=.4]{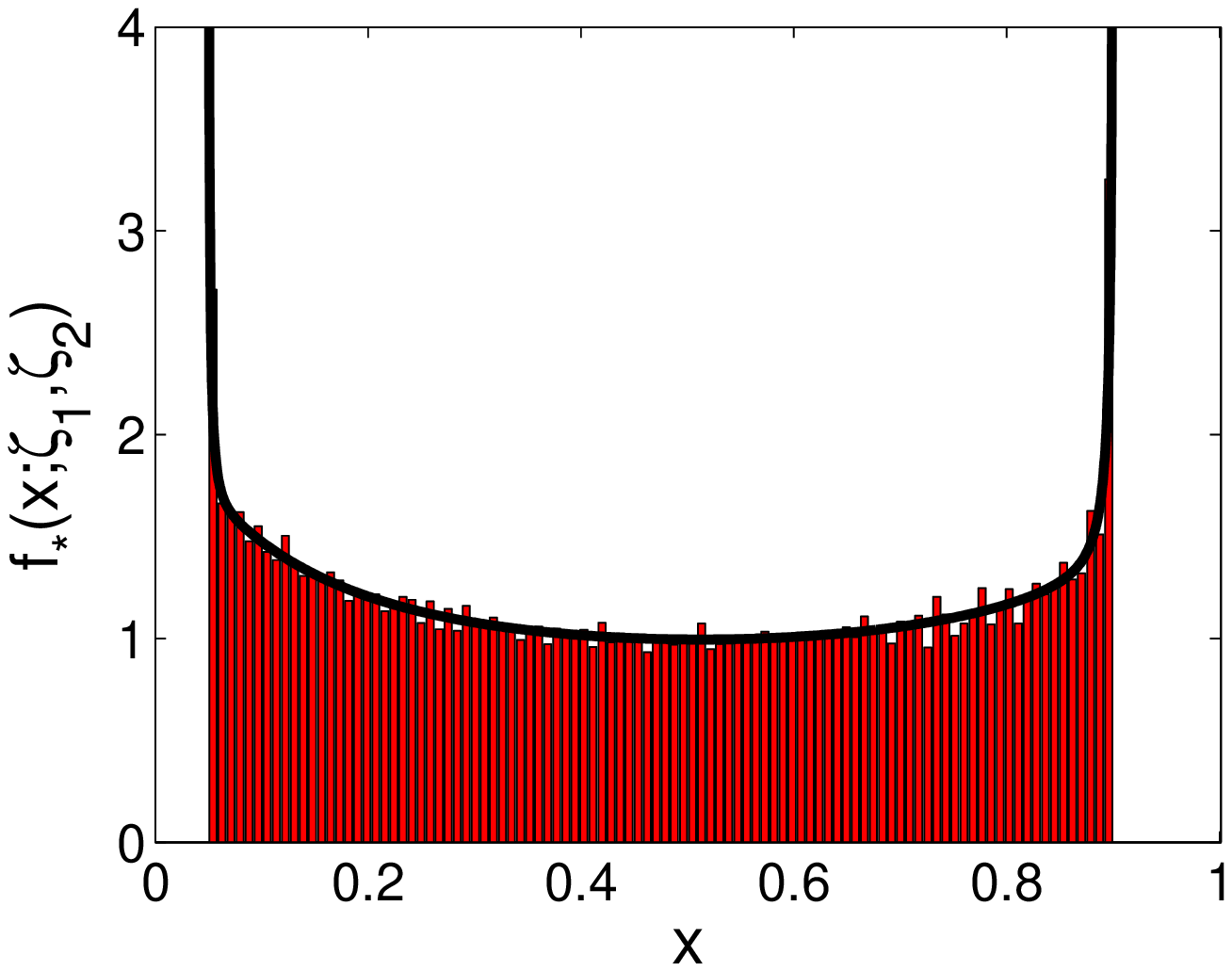}
\label{fig:Ncase4}
}
\label{fig:Nspectrals}
\caption{(Color online) $\beta=2, N=30, \alpha_1=0.6, \alpha_2=0.8$ with $10^5$ samples using the tridiagonal trick. Every Histogram represent the results of the numerical simulation, and the (red) thick line comes from the Coulomb gas prediction given by Eqs.~(\ref{eq:nobarrier}), (\ref{eq:lower}), (\ref{eq:upper}) and (\ref{eq:final}).}
\end{figure}
Next, we compare the Coulomb gas predictions for the probability function $P_N(\lambda_{max}<x)$ and the exact expression of Dumitriu \cite{beta-jacobi}, using matrix hypergeometric functions. For completeness and convenience we quote below the exact result using the notation used in our paper
\begin{eqnarray}
\fl P\left(\lambda_{\max} < x \right) = \frac{{\Gamma_N^{(\beta)} \left(\textstyle{\beta \over 2} N (2+ \aone + \atwo) \right) \Gamma_N^{(\beta)}\left( {{\textstyle{\beta  \over 2}}N + 1 - {\textstyle{\beta  \over 2}}} \right)}}{{\Gamma_N^{(\beta)} \left( {\textstyle{\beta \over 2} N (2+\aone) + 1 - {\textstyle{\beta  \over 2}}} \right)\Gamma_N^{(\beta)}\left( {\textstyle{\beta \over 2} N (1+ \atwo)} \right)}}{x^{{\textstyle{\beta  \over 2}}{N^2}(1+ \aone)}} \times \nonumber \\
\times {}_2F_1^{(\beta)}\left( {{\textstyle{\beta \over 2}}N (1+\aone), - {\textstyle{\beta \over 2}}(N\atwo+ 1);{\textstyle{\beta \over 2}} N (2+\aone) + 1 - {\textstyle{\beta  \over 2}};x{I_N}} \right) \, ,
\label{eq:exact-max}
\end{eqnarray}
where $\Gamma_N^{(\beta)}(c)$ is the multivariate gamma function of parameter $\beta>0$ \cite{beta-jacobi}, ${}_2F_1^{\left( \beta  \right)}\left( {{a_1},{a_2};{b_1};X} \right)$ is the hypergeometric function of matrix argument $X$ and parameter $\beta>0$ \cite{muirhead,beta-jacobi}, and $I_N$ is the $N$-dimensional unit matrix. Ref.~\cite{Dumitriu} gives an alternative expression for the density of the largest eigenvalue, as well as for the smallest eigenvalue. This expression is in practice hard to evaluate, and is not feasible for very large values of $N$. In the following, when considering numerical results, we have used the available MATLAB package by Koev and co-workers \cite{HG1,HG2} to numerically evaluate this matrix hypergeometric function for $N=5$ and $\beta=2$. Figure 7 shows a comparison of the exact result with our predictions. As can be seen the agreement is very good, which implies that $N=5$ is already quite large for the large-$N$ limit to hold in the Coulomb gas results. A few comments are in place here - as mentioned above, larger values of $N$ are already challenging computationally for the matrix hypergeometric function. Furthermore, obtaining precise estimates of the probability density near the hard wall at $\zeta=1$ is very challenging using the exact formula and the MATLAB package, which limits further the applicability of the exact formula.
\begin{figure}[ht]
\centerline{\includegraphics[scale=0.6]{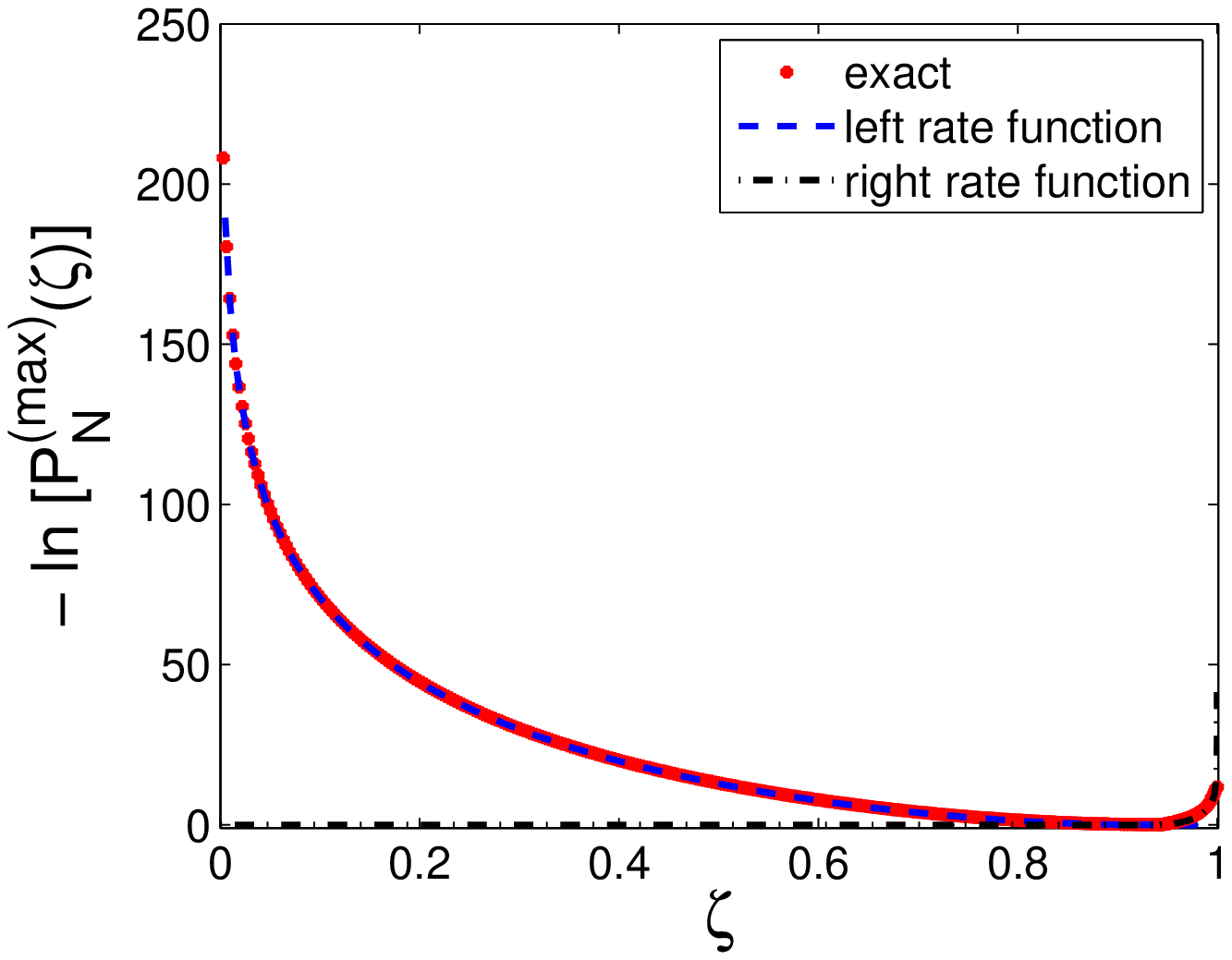}}
\label{fig:rateF}
\caption{(Color online) Results for the largest eigenvalue distribution $-\ln P_N^{(\max)}(\zeta)$ vs $\zeta$. Here $N=5$, $M_1=8$ and $M_2=9$ ($\aone=3/5, \atwo=4/5$), Jacobi matrices are complex ($\beta=2$). The large deviation functions (dashed lines) compare very well with the exact result of \cite{beta-jacobi,Dumitriu} (red dots).}
\end{figure}
To test larger values of $N$ and other values of $\beta$, we have performed simulations based on biased sampling \cite{Nadal11b,Nadal}. We adapt here the nice explanation in \cite{Nadal} to our needs. The Coulomb gas is simulated using the Metropolis algorithm. This works as follows: One proposes a small move $\bm{\lambda}\to\bm{\lambda}'$, which is then accepted  with probability
\begin{equation*}
p={\rm min}\left(\frac{P(\bm{\lambda}')}{P(\bm{\lambda})},1\right)={\rm min}\left(e^{-\frac{\beta}{2}\left(F(\bm{\lambda}')-F(\bm{\lambda})\right) },1\right)
\end{equation*}
or rejected with probability $1-p$.\\
Now we want to study the statistics of the maximum eigenvalue given by $P_{N}^{({\rm max})}(\zeta)$.  This distribution would be peaked around its average, so events in the tails, the ones we are interested in, will be extremely rare. Thus, the standard Metropolis algorithm will provide a very poor sampling of the tails, as explained in \cite{Nadal}.\\
To force the algorithm to explore regions of rare events, we use a trick not dissimilar to its mathematical counterpart: we put a barrier $\lambda^{\star}$ and write $P_{N}^{({\rm max})}(\zeta)$ as
\begin{eqnarray*}
P_{N}^{({\rm max})}(\zeta)=P_{N}^{({\rm max})}(\zeta|\zeta\leq \lambda^{\star})P(\zeta\leq \lambda^{\star})
\end{eqnarray*}
The probability $P_{N}^{({\rm max})}(\zeta|\zeta\leq \lambda^{\star})$ means the probability that the maximum eigenvalue takes the value $\zeta$ conditioned on $\zeta$ being smaller or equal to the barrier $\lambda^{\star}$. This probability can be estimated readily using the Metropolis algorithm. Indeed, we simply start with a configuration such that $\lambda_{{\rm max}}\leq\lambda^\star$ which is then updated according to the Metropolis algorithm, rejecting any move with $\lambda_{{\rm max}}>\lambda^\star$. This allows us to construct histograms of $P_{N}^{({\rm max})}(\zeta|\zeta\leq \lambda^{\star})$ for various values of the barrier $\lambda^\star$. Note that for each value of the barrier $\lambda^\star$ the algorithm only allows us to explore a small region around $\zeta$, that is $\zeta\in[\lambda^\star -\delta, \lambda^\star]$ for some small $\delta$. Finally, since we can write:
\begin{eqnarray*}
\hspace{-2cm}\Phi^{({\rm max})}_N(\zeta)&=&-\frac{\log P_{N}^{({\rm max})}(\zeta|\zeta\leq \lambda^{\star})}{\beta N^2}+K_{\lambda^\star}\,,\quad\quad K_{\lambda^\star}=-\frac{\log P(\zeta\leq \lambda^{\star})}{\beta N^2}\,,
\end{eqnarray*}
where the constant $K_{\lambda^\star}$ depends only on $\lambda^\star$  and $\zeta\in[\lambda^\star -\delta, \lambda^\star]$, we can estimate the derivative of  the rate function $\Phi^{({\rm max})}_N$ at $\zeta$ only from the knowledge of $ P_{N}^{({\rm max})}(\zeta|\zeta\leq \lambda^{\star})$. In other words, full knowledge of $K_{\lambda^\star}$ is not essential to estimate $\Phi'^{(\max)}(\zeta)$.\\
We have implemented this procedure for $\beta=1$, $N=50$, $M_1=80$ and $M_2=90$ (the same $\alpha_i$'s as above) and the estimates for the  the derivative of rate function $\Phi'^{(\max)}(\zeta)$ is depicted in Figure \ref{fig:biased} together with the predictions based on the Coulomb gas approach derived above. As can be seen, for $N=50$ the agreement between the prediction and simulation is very good, and essentially indistinguishable visually. 
\begin{figure}[ht]
\centerline{\includegraphics[scale=0.5]{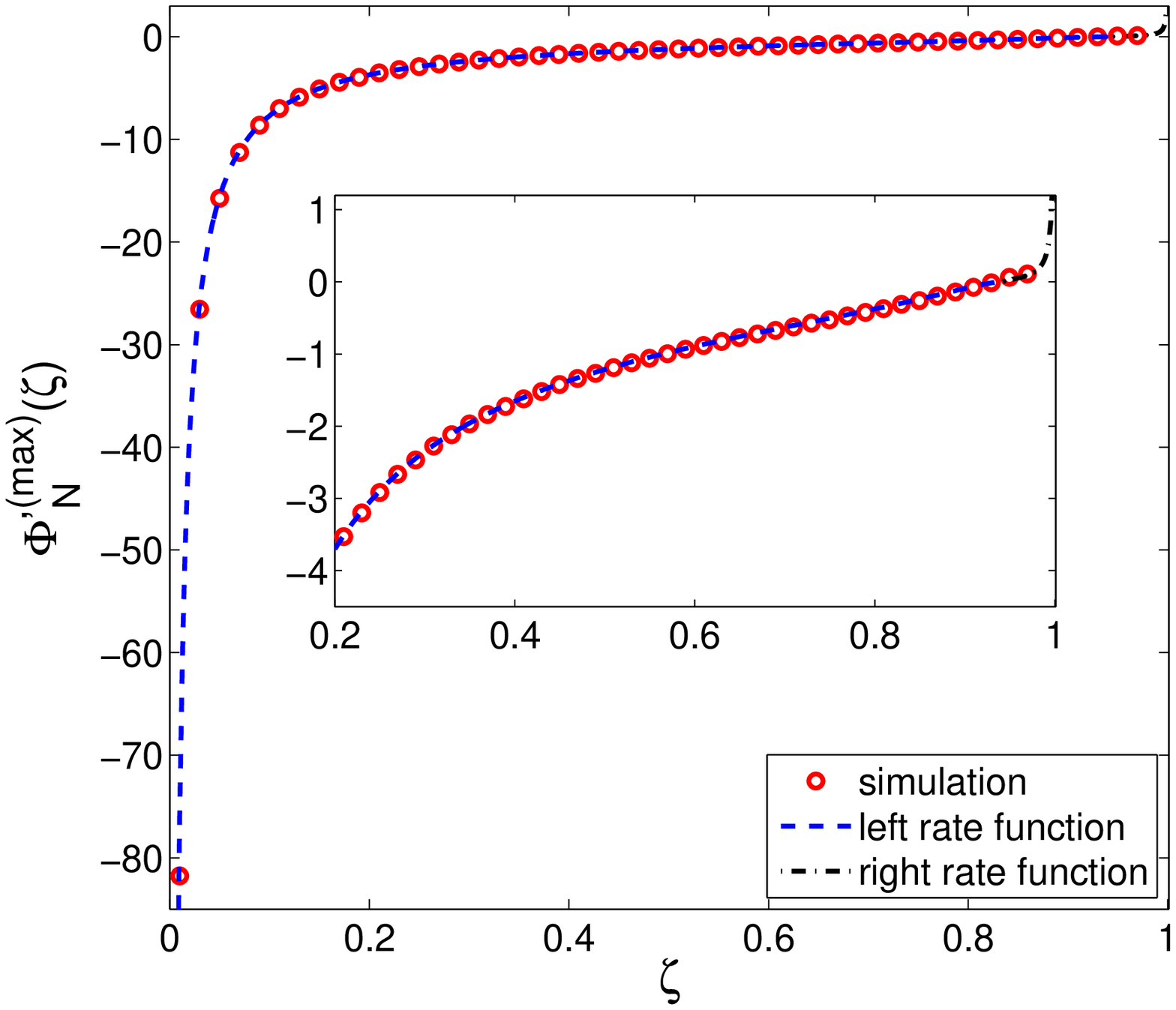}}
\caption{(Color online) Results for the derivative of the large deviation functions of the largest eigenvalue distribution $\Phi'^{(\max)}$ based on simulations. Here $N=50$, $M_1=80$ and $M_2=90$ ($\aone=3/5, \atwo=4/5$), Jacobi matrices are real ($\beta=1$). The derivative of Coulomb gas prediction for the large deviation functions (dotted and dashed lines) compare very well with the simulations (red circles). The inset allows a better resolution in the restricted range.}
\label{fig:biased}
\end{figure}

\section{Summary and Conclusions}
\label{sec:concl}
In this work we study the large deviation functions of the extreme eigenvalue of the Jacobi ensemble of random matrices using the Coulomb gas approach. We obtain explicit expressions for the spectral densities, for the joint probability distribution of the extreme eigenvalues and both the right and left rate functions that describe large fluctuations around the mean value of the extreme eigenvalues. These expressions are expected to hold for large matrices, and a comparison to numerical simulations confirms this. Furthermore, the numerical results as well as the available exact expression suggests than even for fairly small values of $N$ (such as $N=5$) our results offer a reasonable approximation. 
This implies a potential important application of this work, namely providing a readily computable approximation for the probabilities of the extreme eigenvalues -- expressions which may be useful in multivariate statistics. Among the numerous techniques require information about the extreme eigenvalues of the Jacobi ensemble are Canonical Correlation Analysis (CCA), the largest root test for a multiple response linear model, and more -- as discussed earlier in the introduction and in many references in \cite{johnstone,johnstone2,Dumitriu}.

An interesting extension of this work could be the calculation of sub-leading corrections to the various distributions and rate functions calculated in this paper. This is important especially since these corrections round off the sharp boundaries at $\zeta_\pm$ and describe the appearance of tails for any finite $N$. This may be achieved by using a recent method developed for Gaussian matrices, and presented in Refs.~\cite{Borot11,Borot12} where the the corresponding left/right rate functions are calculated. It is hoped that this method could be applied to the Jacobi ensemble.


\section{Acknowledgments}
After submitting this paper, we were made aware by P. Forrester about his recent work~\cite{Forrester12}, where an alternative method is used to derived the ${\vO}(N)$ rate function $\Phi^{(\max)}_+(x)$ (given by Eq.~(\ref{eq:rldv-Jacobi}) above) as well as its ${\vO}(1)$ contribution. We thank him for pointing out his paper to us.

\newpage
\appendix
\section{Typical fluctuation of the extreme eigenvalues in terms of the Tracy--Widom distribution}
\label{app:TW}
It is known \cite{johnstone,johnstone2} that typical deviations of $\lambda_{\max}$ are related to the Tracy--Widom distribution \cite{TW}. In particular, the logit transform of $\lambda_{\max}$, namely
\begin{equation*}
W \equiv  {\rm logit} (\lambda_{\max}) =\log\left( \frac{\lambda_{\max}}{1-\lambda_{\max}} \right) \, ,
\end{equation*}
approaches the Tracy--Widom distribution in the large $N$ limit, after being properly centered and scaled, as follows
\begin{equation*}
\frac{W(N,\aone,\atwo) - \mu_W(N,\aone,\atwo)}{\sigma_W (N,\aone,\atwo)}  \stackrel{N \to \infty}\sim  F_\beta \, ,
\end{equation*}
where $F_\beta$ is the TW distribution with the corresponding Dyson index $\beta$ (proven in \cite{johnstone} for $\beta=1,2$). The centering and scaling parameters are given by
\begin{eqnarray*}
\mu_W (N,\aone,\atwo)      &=&  \log \frac{\zeta_+}{1-\zeta_+}  \, , \\
\sigma_W^3 (N,\aone,\atwo) &=&  \frac{1}{N^2 \sqrt{(1+ \aone)(1+ \atwo)(1+ \aone + \atwo)} \zeta_+ (1-\zeta_+)} \, .
\end{eqnarray*}
Therefore, if $Z_\beta \sim F_\beta$, then $\lambda_{\max} \sim 1/(1+\e^{-\mu_W - \sigma_W z_\beta})$.\\
As explained on in \cite{johnstone} (p. 2651) one can write this result informally for $\lambda_{\max}$ as
\begin{equation*}
\lambda_{\max} \sim \mu_{\max} + N^{-2/3}\sigma_{\max} Z_\beta + \Or \left( N^{-4/3} \right) \, ,
\end{equation*}
where
\begin{eqnarray*}
\mu_{\max}      &=&  \zeta_+  \, , \\
\sigma_{\max}^3 &=&  \frac{\left[ \zeta_+ (1-\zeta_+) \right]^2}{\sqrt{(1+ \aone)(1+ \atwo)(1+ \aone + \atwo)}} \, .
\end{eqnarray*}
Note that this result requires $\aone,\atwo>0$, so that the case of square or almost square matrices is not described by TW.

By symmetry, one can show that the distribution of $\lambda_{\min}(N,\aone,\atwo)$ is the same as that of $1-\lambda_{\max}(N,\atwo,\aone)$ (note the exchange of $\aone$ and $\atwo$). Therefore, one can easily get the typical fluctuations of $\lambda_{\min}$ from the results presented above.

\include{transapp}

\include{integralsapp}

\section{Useful identities for $\zeta_-$ and $\zeta_+$}
\label{app:identities}
\begin{eqnarray*}
\sqrt{\zeta_+ \zeta_-} &=& \frac{\aone}{2 + \aone + \atwo} \\
\sqrt{(1 - \zeta_+)(1 - \zeta_-)}  &=& \frac{\atwo}{2 + \aone + \atwo} \, , \\
\Delta_- = \zeta_+ - \zeta_- &=& \frac{4{\sqrt{(1+ \aone)(1+ \atwo)(1+ \aone + \atwo)} }}{{{{(2 + \aone + \atwo)}^2}}} \, , \\
\zeta_+ + \zeta_- &=& 2\frac{{\alpha _1^2 + 2 + 2\atwo + 2\aone + \aone\atwo}}{{{{\left( {2 + \aone + \atwo} \right)}^2}}} \, , \\
\sqrt{\zeta_+}  - \sqrt{\zeta_-} &=& \frac{2\sqrt{1 + \atwo}}{2 + \aone + \atwo} \, , \\
\sqrt{\zeta_+}  + \sqrt{\zeta_-} &=& \frac{2\sqrt{\left( {1 + \aone} \right)(1 + \aone + \atwo)}}{2 + \aone + \atwo} \, , \\
\sqrt{1 - \zeta_+}  - \sqrt{1 - \zeta_-} &=&  - \frac{2\sqrt{1 + \aone}}{2 + \aone + \atwo} \, , \\
\sqrt{1 - \zeta_+}  + \sqrt{1 - \zeta_-} &=& \frac{2\sqrt{(1 + \atwo)(1 + \aone + \atwo)}}{2 + \aone + \atwo} \, , \\
\sqrt{\zeta_+(1 - \zeta_-)} + \sqrt{\zeta_-(1 - \zeta_+)} &=& \frac{2\sqrt{(1 + \atwo)(1 + \aone)}}{2 + \aone + \atwo} \, , \\
\sqrt{\zeta_+ (1 - \zeta_-)} - \sqrt{\zeta_-(1 - \zeta_+)} &=& \frac{2\sqrt{1 + \atwo + \aone}}{2 + \aone + \atwo} \, .
\end{eqnarray*}

\newpage

\bibliographystyle{unsrt}
\bibliography{bibfile}

\end{document}

%% file: transapp.tex
\section{Transformation of Integrals}
\label{app:transformations}

The Tricomi solution of the integral equation given by Eq.~\eref{eq:tricomi} can be expanded as
\begin{equation}
f_*(x)=\frac{1}{\pi^2}\frac{ -\frac{\alpha_1}{2}P\int_{\zone}^{\ztwo}\frac{\sqrt{(t-\zone)(\ztwo-t)}}{t(t-x)}dt -
\frac{\alpha_2}{2}P\int_{\zone}^{\ztwo}\frac{\sqrt{(t-\zone)(\ztwo-t)}}{(1-t)(t-x)}dt + C }{\sqrt{(x-\zone)(\ztwo-x)}} \, .
\label{eq:pfint}
\end{equation}
We can further factorize the first integral from \eref{eq:pfint} into partial fractions
\begin{equation*}
\fl \int_{\zone}^{\ztwo}\frac{\sqrt{(t-\zone)(\ztwo-t)}}{t(t-x)}dt
=\frac{1}{x}\int_{\zone}^{\ztwo}\frac{\sqrt{(t-\zone)(\ztwo-t)}}{t-x}dt
-\frac{1}{x}\int_{\zone}^{\ztwo}\frac{\sqrt{(t-\zone)(\ztwo-t)}}{t}dt
\end{equation*}
Let's start with the first integral in the last expression, where we can use the following change of variables: $t=\zone+y \Delta$, $x=\zone+c \Delta$, where $\Delta=\ztwo-\zone$
\begin{equation*}
P \int_{\zone}^{\ztwo}\frac{\sqrt{(t-\zone)(\ztwo-t)}}{t-x}dt = \xD \cdot P \int_0^1\frac{\sqrt{y(1-y)}}{y-c}dy \, .
\end{equation*}
We can use the following integral
\begin{equation}
P\int_0^1\frac{\sqrt{y(1-y)}}{y-c}dy = \pi\left(\frac{1}{2}-c\right) \, ,
\label{eq:result}
\end{equation}
where $0<c<1$, and which leads to
\begin{equation}
\int_{\zone}^{\ztwo}\frac{\sqrt{(t-\zone)(\ztwo-t)}}{t-x}dt \quad =
\pi\left(\frac{\ztwo+\zone-2x}{2}\right)
\label{eq:second}
\end{equation}
Similarly, 
\begin{equation*}
\int_{\zone}^{\ztwo}\frac{\sqrt{(t-\zone)(\ztwo-t)}}{t}dt
=\xD\int_0^1\frac{\sqrt{t(1-t)}}{t+\zone/\xD}dt
=\frac{\pi}{2}\left(\sqrt{\ztwo}-\sqrt{\zone}\right)^2 \, ,
\end{equation*}
where we have used
\begin{equation}
\int_0^1\frac{\sqrt{t(1-t)}}{t+c}dt = \frac{\pi}{2}\left(\sqrt{c+1}-\sqrt{c}\right)^2 \, .
\label{eq:first}
\end{equation}
Using results \eref{eq:second} and \eref{eq:first} we can write the above as
\begin{equation*}
\int_{\zone}^{\ztwo}\frac{\sqrt{(t-\zone)(\ztwo-t)}}{t(t-x)}dt
=\frac{\pi}{x}\left(\sqrt{\zone\ztwo} - x\right)
\end{equation*}
The second integral of \eref{eq:pfint} can be factorized as
\begin{equation*}
\fl \int_{\zone}^{\ztwo}\frac{\sqrt{(t-\zone)(\ztwo-t)}}{(1-t)(t-x)}dt
=\frac{1}{1-x}\int_{\zone}^{\ztwo}\frac{\sqrt{(t-\zone)(\ztwo-t)}}{1-t}dt
+\frac{1}{1-x}\int_{\zone}^{\ztwo}\frac{\sqrt{(t-\zone)(\ztwo-t)}}{(t-x)}dt
\end{equation*}
The second integral in the last expression is the same as before and so we are left with the first integral, in which we use the same methods as before to get
\begin{equation*}
\int_{\zone}^{\ztwo}\frac{\sqrt{(t-\zone)(\ztwo-t)}}{1-t}dt = \frac{\pi}{2}\left(\sqrt{1-\ztwo}-\sqrt{1-\zone}\right)^2
\end{equation*}
Gathering results, we can finally write the following simplified expression for $f(x;\zone,\ztwo)$
\begin{equation*}
\fl f_*(x;\zone,\ztwo) =\frac{1}{2\pi}\frac{1}{\sqrt{(x-\zone)(\ztwo-x)}}\left[\frac{\alpha_1}{x}\left(x-\sqrt{\zone\ztwo} \right)
+\frac{\alpha_2}{1-x}\left(1-x-\sqrt{(1-\zone)(1-\ztwo)}\right)+C\right]
\end{equation*}
Since we need a normalized expression for $f_*(x)$ i.e. $\int_{\zone}^{\ztwo}f_*(x)dx = 1$ (recall that this was enforced using the Lagrange multiplier $A$), this determines $C=2$, and thus we have the final expression for the limiting spectral density, namely
\begin{equation*}
\fl f_*(x;\zone,\ztwo)=\frac{1}{2\pi}\frac{1}{\sqrt{(x-\zone)(\ztwo-x)}}\left[\alpha_1\left(1-\frac{\sqrt{\zone\ztwo}}{x}\right)
+\alpha_2\left(1-\frac{\sqrt{(1-\zone)(1-\ztwo)}}{1-x}\right)+2\right] \, .
\end{equation*}

%% file: integralsapp.tex
\section{Useful Integrals}
\label{app:case1}
In this appendix we provide details regarding certain integrals which are needed in various derivations described in the text. This is done for convenience since some of these expressions are not easily available in a single table such as \cite{rhyz}. Some formulas are a result of a non-trivial simplification of expressions involving  hypergeometric functions, and some integrals were performed by the authors.

Integrals $\vI_1-\vI_9$ are used to obtain the simple expression in \Eref{eq:finalaction} and the following shows how this is achieved. After determining the multiplier $A$
\begin{equation*}
\fl A = 2 \alpha_1\log \left(\frac{\sqrt{\zeta_1} + \sqrt{\zeta_2}}{2}\right)
+2\alpha_2\log\left(\frac{\sqrt{1-\zeta_1}+\sqrt{1-\zeta_2}}{2}\right)
+2\log\frac{\xD}{4} \, ,
\end{equation*}
where $\xD=\zeta_2-\zeta_1$.Inserting this into Eq.~\eref{eq:action}, we find that $S[f_*(x);\zeta_1,\zeta_2]$ is given by:
\begin{equation*}
\fl S[f_*(x);\zeta_1,\zeta_2] =-\frac{A}{2}-\frac{\alpha_1}{2}\int_{\zeta_1}^{\zeta_2}dx f_*(x)\log x
-\frac{\alpha_2}{2}\int_{\zeta_1}^{\zeta_2}dx f_*(x)\log(1-x) \, ,
\end{equation*}
which can also be expressed as
\begin{eqnarray*}
\fl S[a,c,\Delta] =-\frac{A}{2}-\frac{\alpha_1}{2}\left[(\alpha_1+\alpha_2+2)\left(\log\Delta\,\vI_1+\vI_2(c)\right)
+\alpha_1\sqrt{\zeta_1\zeta_2}\left(\frac{\log\Delta}{\Delta}\,\vI_3(c) +\frac{1}{\Delta}\,\vI_4(c)\right) \right. \\
\fl \left. + \alpha_2\sqrt{(1-\zeta_1)(1-\zeta_2)}\left(\frac{\log\Delta}{\Delta}\,\vI_5(a) +\frac{1}{\Delta}\,\vI_6(a,c)\right)\right] 
-\frac{\alpha_2}{2}\left[(\alpha_1+\alpha_2+2)\left(\log\Delta\,\vI_1+\vI_7(a)\right) \right.\\
\fl \left. +\alpha_1\sqrt{\zeta_1\zeta_2}\left(\frac{\log\Delta}{\Delta}\,\vI_3(c)+\frac{1}{\Delta}\,\vI_8(a,c)\right)
+\alpha_2\sqrt{(1-\zeta_1)(1-\zeta_2)}\left(\frac{\log\Delta}{\Delta}\,\vI_5(a)+\frac{1}{\Delta}\,\vI_9(a)\right)\right]
\end{eqnarray*}
where $c=\frac{\zeta_1}{\Delta}, a=\frac{1-\zeta_1}{\Delta}$, and with $c>0$ and $a>1$.
After making the change the variables as in Appendix~\ref{app:transformations} of $x=y\xD+\zeta_1$, the integrals $\vI_k$ are given by the following formulas

\begin{eqnarray*}
\fl\vI_1 &=& \int_0^1 \frac{1}{\sqrt{y(1-y)}}\,dy =\pi\\
\fl\vI_2(c) &=& \int_0^1\frac{\log\left(y+c\right)}{\sqrt{y(1-y)}}\,dy = 2
\pi  \log \left(\frac{\sqrt{c}+\sqrt{c+1}}{2} \right)\\
\fl\vI_3(c) &=& \int_0^1\frac{1}{\left(y+c\right)\sqrt{y(1-y)}}\,dy
=\frac{\pi}{\sqrt{c(c+1)}}\\
\fl\vI_4(c) &=& \int_0^1\frac{\log\left(y+c\right)}{\left(y+c\right)\sqrt{y(1-y)}}\,dy = \frac{\pi}{\sqrt{c(c+1)}}\log\left[4c(c+1)\left(\sqrt{c+1}-\sqrt{c}\right)^2\right]\\
\fl\vI_5(a) &=& \int_0^1\frac{1}{\left(a-y\right)\sqrt{y(1-y)}}dy =\frac{\pi}{\sqrt{a(a-1)}}\\
\fl\vI_6(a,c) &=& \int_0^1\frac{\log\left(y+c\right)}{\left(a-y\right)\sqrt{y(1-y)}}\,dy \nonumber \\
\fl &=& \frac{2 \pi}{\sqrt{a (a-1)}} \left[\log \left(\sqrt{a(c+1)}+\sqrt{(a-1) c}\right)+\log\left(\sqrt{a}-\sqrt{a-1}\right)\right] \\
\fl\vI_7(a) &=& \int_0^1\frac{\log\left(a-y\right)}{\sqrt{y(1-y)}}\,dy =2
\pi\log\left(\frac{\sqrt{a}+\sqrt{a-1}}{2}\right)\\
\fl\vI_8(a,c)&=& \int_0^1\frac{\log\left(a-y\right)}{\left(y+c\right)\sqrt{y(1-y)}}\,dy \nonumber \\
\fl &=& \frac{2 \pi}{\sqrt{c (c+1)}} \biggl[\log \left(\sqrt{a(c+1)}+\sqrt{(a-1) c}\right)+\log\left(\sqrt{c+1}-\sqrt{c}\right)\biggr]\\
\fl\vI_9(a) &=& \int_0^1\frac{\log\left(a-y\right)}{\left(a-y\right)\sqrt{y(1-y)}}\,dy \nonumber \\
\fl &=& \frac{\pi}{\sqrt{a(a-1)}}\log\left[4a(a-1)\left(\sqrt{a}-\sqrt{a-1}\right)^2\right] =\vI_4(a-1)
\end{eqnarray*}
Gathering these results and after some simple algebra, we eventually get the final expression of $S[f_*(x);\zeta_1,\zeta_2]$ as in equation \eref{eq:finalaction}.
\begin{eqnarray*}
\fl S[f_*(x);\zone,\ztwo] &=& -(\aone+\atwo+2) \left(\aone\log\frac{\sqrt{\zone}+\sqrt{\ztwo}}{2}
+\atwo\log\frac{\sqrt{1-\zone}+\sqrt{1-\ztwo}}{2} \right) \nonumber \\
\fl &+& \frac{{\aone}^2}{2} \log \sqrt{\zone\ztwo}+\frac{{\atwo}^2}{2} \log \sqrt{(1-\zone)(1-\ztwo)} \nonumber \\
\fl &+&\aone\atwo \log\frac{\sqrt{\ztwo(1-\zone)}+\sqrt{\zone(1-\ztwo)}}{2}-\log\frac{\ztwo-\zone}{4}
\end{eqnarray*}

To derive the simple expression given in \Eref{eq:rldv-Jacobi} we need the following two integrals
\begin{eqnarray}
\fl \vI_{10}(x,a) &=& \int\limits_0^1{\frac{{\sqrt{t\left({1 - t}\right)} \ln \left({x-t}\right)}}{{t + a}}dt}
= \frac{\pi}{2}\left[{{\left({\sqrt x-\sqrt{x-1}}\right)}^2} + 2\left({2a+1}\right)\ln \left({\frac{{\sqrt x +\sqrt{x-1}}}{2}}\right)\right. \nonumber\\
\fl &-& \left. 4\sqrt{a\left({a+1}\right)} \ln\left({\frac{{\sqrt{\left({a+1}\right)x} + \sqrt{a\left({x-1}\right)}}}{{\sqrt{a+1} +\sqrt a}}} \right) \right]
\label{eq:J1}
\end{eqnarray}
where $x>1$ and $a>0$, and
\begin{eqnarray}
\fl \vI_{11}(x,a) &=& \int\limits_0^1 {\frac{{\sqrt{t\left({1-t}\right)} \ln \left({x-t}\right)}}{{t-a}}dt}
= \frac{\pi}{2}\left[{{\left({\sqrt x-\sqrt{x-1}}\right)}^2} -2\left({2a-1} \right)\ln \left({\frac{{\sqrt x+\sqrt{x-1}}}{2}}\right)\right. \nonumber \\
\fl &+& \left. 4\sqrt{a\left({a-1}\right)} \ln \left({\frac{{\sqrt{\left({a-1}\right)x}+\sqrt{a\left({x-1}\right)}}}{{\sqrt {a-1}+\sqrt a}}}\right)\right]
\label{eq:J2}
\end{eqnarray}
where $x>1$ and $a>1$.

To achieve the simple expression given in \Eref{eq:lldv-Jacobi} we need the following two integrals
\begin{eqnarray}
\fl \vI_{12}(x,a) &=& \int\limits_0^1 {\frac{{\sqrt {t\left( {1 - t}\right)} \ln \left( {t + x} \right)}}{{t + a}}dt}
= \frac{\pi}{2}\left[ { - {{\left( {\sqrt {x + 1}  - \sqrt x } \right)}^2} +2\left( {1 + 2a} \right)\ln \left( {\frac{{\sqrt x  + \sqrt {x + 1}
}}{2}} \right)}\right. \nonumber \\
\fl &-& \left.4\sqrt {a\left( {a + 1} \right)} \ln \left( {\frac{{\sqrt {\left( {a + 1} \right)x}  + \sqrt {a\left( {x + 1} \right)} }}{{\sqrt {a + 1}
+\sqrt a }}} \right) \right]
\label{eq:J3}
\end{eqnarray}
where $x>0$ and $a>0$, and
\begin{eqnarray}
\fl \vI_{13}(x,a) &=& \int\limits_0^1 {\frac{\sqrt{t(1 - t)} \ln(t+x)}{t-a}dt}
=\frac{\pi }{2}\left[ -\left( {\sqrt {x + 1}  - \sqrt x } \right)^2 + 2(1-2a) \ln \left(\frac{\sqrt{x}  + \sqrt{x+1}}{2}\right) \right. \nonumber \\
\fl &+& \left. 4 \sqrt{a(a-1)} \ln \left(\frac{\sqrt{(a-1)x} + \sqrt{a(x+1)}}{\sqrt{a-1} + \sqrt{a}} \right) \right]
\label{eq:J4}
\end{eqnarray}
where $x>0$ and $a>1$.